\documentclass[12pt,thmsa]{article}
\usepackage{sw20lart}

\input{tcilatex}
\begin{document}

\author{Emilio Santos}
\title{Gravity effects of the quantum vacuum. Dark energy and dark matter}
\date{May, 11, 2015}
\maketitle
\tableofcontents

\begin{abstract}
The stress-energy tensor of the quantum vacuum is studied for the particular
case of quantum electrodynamics (QED), that is a fictituous universe where
only the electromagnetic and the electron-positron fields exist. The
integrals involved are ultraviolet divergent but it is suggested that a
natural cut-off may exist. It is shown that, in spite of the fact that the
stress-energy tensor of the electromagnetic field alone is traceless (i.e
the pressure P equals 1/3 the energy density u), the total QED tensor is
proportional to the metric tensor to a good approximation (i. e. P = -u). It
is proposed that there is a cosmological constant in Einstein equation that
exactly balances the stress-energy of the vacuum. It is shown that vacuum
fluctuations give rise to a modified spacetime metric able to explain dark
energy. Particular excitations of the vacuum are studied that might explain
dark matter.
\end{abstract}

\section{ Introduction}

\subsection{Quantum vacuum, cosmological constant and dark energy}

Quantum field theory predicts that the vacuum is not empty, but filled with
interacting quantum fields. A calculation of the energy of the fields gives
divergent quantities, positive for free Bose fields and negative for free
Fermi fields, as will be illustrated in section 2 below for the particular
case of quantum electrodynamics (QED). Nevertheless the energy of the vacuum
is irrelevant in most quantum calculations not including gravity. In fact it
may be ignored defining the zero of energy at the level of the vacuum. In
practice this is achieved using the ``normal ordering rule'' for the
creation and annihilation field operators.

However a big difficulty appears if we want to take the vacuum gravity into
account, because in this case the zero of energy cannot be fixed at will.
Two alternative solutions have been proposed, none of them completely
satisfactory: 1) Assuming that the vacuum fields represent a mathematical
artefact of the quantum formalism and only the excitations of the fields
above the vacuum contribute actual energy, 2) The vacuum fields have energy,
but positive and negative contributions may cancel to each other, at least
approximately. In this paper the first possibility is rejected and the
second one is studied searching for clues for the solution of a number of
well known difficulties.

If there is an actual energy density of the vacuum there should be also
pressure terms, that is a full stress-energy tensor. Indeed if the spacetime
background is Minkowski, at least approximately, then it seems that the
vacuum properties should be Lorentz invariant whence the vacuum
stress-energy tensor would be proportional to the metric tensor. As a
consequence the effect of that tensor would be equivalent to a cosmological
term in the Einstein equation of general relativity. The existence of such a
term might explain the observations in cosmology, which have shown that the
universe is in accelerated expansion\cite{Peebles}\cite{Sahni}. Actually the
cause of that expansion is unknown and it is usually named ``dark energy'',
but it is common wisdom to interpret dark energy as an effect of the quantum
vacuum. Other alternatives have been proposed that will not be commented
here.

The hypothesis that dark energy is an effect of the quantum vacuum poses a
well known problem\cite{Weinberg}. In fact it is plausible that the vacuum
energy density, $\rho _{DE},$ should correspond to a combination of the
universal constants $c,$ $%TCIMACRO{\UNICODE[m]{0x127}}
%BeginExpansion
\rlap{\protect\rule[1.1ex]{.325em}{.1ex}}h%
%EndExpansion
,G,$ that is 
\begin{equation}
\rho _{DE}\sim \frac{c^{5}}{G^{2}%TCIMACRO{\UNICODE[m]{0x127}}
%BeginExpansion
\rlap{\protect\rule[1.1ex]{.325em}{.1ex}}h%
%EndExpansion
}\simeq 10^{97}kg/m^{3}.  \label{Planckdensity}
\end{equation}
This is Planck's density whose value is about 123 orders greater than the
required dark energy density, with current empirical value\cite{Planck}

\begin{equation}
\rho _{DE}=-P_{DE}\simeq (6.0\pm 0.2)\times 10^{-27}kg/m^{3},\smallskip
\label{darkdens}
\end{equation}
$P_{DE}<0$ being the pressure. The first eq.$\left( \ref{darkdens}\right) $
is consistent with the vacuum stress-energy tensor being Lorentz invariant.

\subsection{The relevance of the vacuum fluctuations}

In this paper a solution is proposed to the problem of the big disagreement
between the ``theoretical'' value eq.$\left( \ref{Planckdensity}\right) $,
and the empirical result eq.$\left( \ref{darkdens}\right) $. In order to
understand the logic of the proposal I start pointing out the relevance of
the fluctuations of the vacuum fields, that I believe has not been fully
appreciated in the analysis of the gravitational effects of the vacuum. For
the sake of clarity I will begin with arguments involving classical (rather
than quantum) fields and Newtonian gravity (rather than general relativity).

Let us assume that the vacuum consists of a stationary, homogeneous and
isotropic set of interacting fields, some of the fields contributing
positive energy and other fields negative energy and similarly for the
interactions. If we want to escape from the absurd assumption that the field
energy density is at the Planck scale (i. e. eq.$\left( \ref{Planckdensity}%
\right) )$ a plausible hypothesis is that positive and negative
contributions cancel to each other. A partial cancelation giving the result
eq.$\left( \ref{darkdens}\right) ,$ that is more than hundred orders smaller
than Planck's density, looks conspiratory\cite{Weinberg}. Therefore it is
plausible that the cancelation is exact, which I will assume at this moment
(this assumption will be later modified, taking quantum theory and general
relativity into account, see section 4 below). If this is the case it seems
that dark energy remains unexplained. However I will argue in the following
that, even if the average energy density of the vacuum is zero, the
fluctuations might explain dark energy.

In fact let us suppose that the vacuum fields fluctuate. That is, at a given
time there are regions with energy density above the mean (that is positive
fluctuations) and other regions with density below it (negative
fluctuations). Thus, assuming that the mean vacuum energy is zero, the
fluctuation of the density would be a function, $\rho \left( \mathbf{r,}%
t\right) ,$ fufilling the condition that its space average is zero, that is 
\begin{equation}
\lim_{V\rightarrow \infty }\int_{V}\rho \left( \mathbf{r,}t\right) d^{3}%
\mathbf{r}=0.  \label{0.3}
\end{equation}
We must assume, as in special relativity, that energy gravitates. Then we
should accept that Newtonian gravity associates negative (positive)
gravitational potential to positive (negative) mass-energy. (Of course
Newtonian gravity is not consistent with special relativity, but for a
simplified argument I may combine both). Thus, using Newtonian theory, the
gravitational energy, $E_{G},$ of the vacuum fluctuations in the volume $V$
would be 
\begin{eqnarray}
E_{G} &=&\frac{1}{2}\int_{V}\rho \left( \mathbf{r}_{1}\mathbf{,}t\right)
\Phi \left( \mathbf{r}_{1}\mathbf{,}t\right) d^{3}\mathbf{r}_{1},  \label{EG}
\\
\Phi \left( \mathbf{r}_{1}\mathbf{,}t\right) &\equiv &-G\int_{V}\frac{\rho
\left( \mathbf{r}_{2}\mathbf{,}t\right) }{\left| \mathbf{r}_{2}-\mathbf{r}%
_{1}\right| }d^{3}\mathbf{r}_{2},  \nonumber
\end{eqnarray}
where $G$ is Newton's constant and $V$ a volume large in comparison with the
range of the fluctuations. As the fluctuating density $\rho $ has zero mean,
the gravitational potential of the fluctuations $\Phi ,$ which is linearly
related to $\rho ,$ is also nil on the average. Thus the vacuum fluctuations
would produce no average gravitational field at any spacetime point. This
might suggest that vacuum fluctuations would not lead to spacetime curvature
when we pass from Newton\'{}s to Einstein\'{}s gravity. If this is the case
the density fluctuations could not explain dark energy. However the\
suggestion\ is\ flawed\ due\ to\ the\ fact\ that\ the\ equations\ of\
Einstein's\ theory\ are\ nonlinear, in contrast with Newton\'{}s, a point
that will be explained in more detail in section 4 below.

There is another argument supporting that vacuum fluctuations, with zero
mean, may produce a long range effect. In fact, the gravitational energy $%
E_{G}$ due to the fluctuations, eq.$\left( \ref{EG}\right) $, is not linear
but quadratic in the fluctuating density. Therefore the mean gravitational
energy density, $\bar{\rho}_{G},$ needs not be zero.

The quantity $\bar{\rho}_{G}$ is defined by 
\begin{eqnarray*}
\bar{\rho}_{G} &=&\lim_{V\rightarrow \infty }\frac{E_{G}}{V} \\
&=&-\lim_{V\rightarrow \infty }\left[ \frac{G}{2V}\int_{V}\rho \left( 
\mathbf{r}_{1}\mathbf{,}t\right) d^{3}\mathbf{r}_{1}\int_{V}\left| \mathbf{r}%
_{2}-\mathbf{r}_{1}\right| ^{-1}\rho \left( \mathbf{r}_{2}\mathbf{,}t\right)
d^{3}\mathbf{r}_{2}\right] .
\end{eqnarray*}
Changing from the variables $\left\{ \mathbf{r}_{1},\mathbf{r}_{2}\right\} $
to the new ones $\left\{ \mathbf{r}_{0},\mathbf{r}\right\} $ via 
\[
\mathbf{r}_{0}=\frac{\mathbf{r}_{1}+\mathbf{r}_{2}}{2},\mathbf{r}=\mathbf{r}%
_{2}-\mathbf{r}_{1}, 
\]
we get 
\begin{eqnarray}
\bar{\rho}_{G} &=&-\frac{G}{2}\int \frac{1}{r}C\left( r\right) d^{3}\mathbf{%
r=}-2\pi G\int_{0}^{\infty }C\left( r\right) rdr,  \label{0.1} \\
C\left( r\right) &\equiv &\lim_{V\rightarrow \infty }\left[ \frac{1}{V}%
\int_{V}\rho \left( \mathbf{r}_{0}\mathbf{+r/}2,t\right) \rho \left( \mathbf{%
r}_{0}-\mathbf{r/}2,t\right) d^{3}\mathbf{r}_{0}\right] .  \nonumber
\end{eqnarray}
We see that the mean graviational energy of the fluctuations depends on the
two-point energy density correlation, $C\left( r\right) $, and it may be
different from zero. The two-point correlation at equal times should depend
only on the distance $r=\left| \mathbf{r}_{2}-\mathbf{r}_{1}\right| ,$ as is
consistent with the assumed homogeneity and isotropy of the vacuum. Also it
fulfils the condition 
\begin{eqnarray*}
\int C\left( r\right) d^{3}\mathbf{r} &=&\lim_{V\rightarrow \infty }\left[ 
\frac{1}{V}\int_{V}d^{3}\mathbf{r}\int_{V}\rho \left( \mathbf{r}_{0}\mathbf{%
+r/}2,t\right) \rho \left( \mathbf{r}_{0}-\mathbf{r/}2,t\right) d^{3}\mathbf{%
r}_{0}\right] = \\
&=&\lim_{V\rightarrow \infty }\left[ \frac{1}{V}\int_{V}\rho \left( \mathbf{r%
}_{1}\mathbf{,}t\right) d^{3}\mathbf{r}_{1}\times \int_{V}\rho \left( 
\mathbf{r}_{2}\mathbf{,}t\right) d^{3}\mathbf{r}_{2}\right] =0,
\end{eqnarray*}
taking eq.$\left( \ref{0.3}\right) $ into account.

The conclusion from our classical analysis is that, in addition to the mean
energy density of the vacuum, if any, there should be a gravitational energy
density caused by the density fluctuations. In section 3 I will calculate
the two-point correlation in quantum electrodynamics. In section 4 the
relation of this correlation with the dark energy density $\rho _{DE},$ eq.$%
\left( \ref{darkdens}\right) ,$ will be studied. Indeed we shall see that $%
\bar{\rho}_{G}$ may be taken as an estimate for $\rho _{DE}.$

The suggestion that quantum vacuum fluctuations may give rise to an
effective cosmological constant was made by Zeldovich\cite{Zel} in 1976. The
possibility that these fluctuations are at the origin of the dark energy has
been explored recently\cite{Santos} using a simplified model involving
general relativity.

\section{Stress-energy tensor of the vacuum fields}

\subsection{Energy density and pressure. Mean and two-point correlations}

Our aim is to study the gravitational effects of the vacuum. According to
general relativity the quantity relevant for that purpose is the
stress-energy tensor. If we assume that space-time is Minkowski (which is a
good approximation for our present purposes, but see section 3) then
stationarity, isotropy and homogeneity of the vacuum lead to a metric tensor
defined by just two quantities, the energy density $\rho $ and the pressure $%
P$. That is the metric tensor should be 
\[
T_{1}^{1}=T_{2}^{2}=T_{3}^{3}=P,T_{0}^{0}=-\rho ,T_{\mu }^{\nu }=0\text{ if }%
\mu \neq \nu . 
\]
If in addition we assume Lorentz invariance, then we should have $P=-\rho $.
In this case the stress-energy tensor is proportional to the metric tensor.
Actually I shall not assume Lorentz invariance from the start, but should
derive it as a result of the calculations. In order to obtain the
gravitational contribution of the vacuum it is enough to get the mean energy
density, the mean pressure and the relevant two-point correlation functions.
The average energy and pressure will be calculated in section 2 and some of
the correlations in section 3.

In the quantum context vacuum expectations of the appropriate operators
should be substituted for averages.The expectation of the energy density is

\begin{equation}
\rho =\left\langle vac\left| \hat{\rho}\left( \mathbf{r},t\right) \right|
vac\right\rangle .  \label{1.1}
\end{equation}
where the operator $\hat{\rho}\left( \mathbf{r},t\right) $ is the quantized
Hamiltonian density, which should include all quantum fields and their
interactions, excluding the gravitational one. In this paper the
gravitational field (that is the spacetime curvature) is considered as
radically different from the remaining fields of nature, although it should
be quantized in some form (see below, section 3.2). In the Heisenberg
picture the Hamiltonian (or energy) density operator may depend on position
and time, but the state should not. The expectation $\rho ,$ eq.$\left( \ref
{1.1}\right) ,$ cannot depend on $\left( \mathbf{r},t\right) $ due to the
traslational and rotational invariance of the vacuum.

The two-point density correlation is the most relevant quantity for the
study of the quantum fluctuations. It is defined in quantum field theory by 
\begin{equation}
C\left( \mathbf{r}_{1},t_{1};\mathbf{r}_{2},t_{2}\right) =\left\langle
vac\left| \hat{\rho}\left( \mathbf{r}_{1},t_{1}\right) \hat{\rho}\left( 
\mathbf{r}_{2},t_{2}\right) \right| vac\right\rangle ,  \label{1.2}
\end{equation}
provided that $\hat{\rho}\left( \mathbf{r}_{1},t_{1}\right) $ commutes with $%
\hat{\rho}\left( \mathbf{r}_{2},t_{2}\right) .$ If they do not commute the
correlation so defined might not be real, having an imaginary part. If this
is the case we will define the correlation in the form 
\begin{equation}
C=\frac{1}{2}\left\langle vac\left| \hat{\rho}\left( \mathbf{r}%
_{1},t_{1}\right) \hat{\rho}\left( \mathbf{r}_{2},t_{2}\right) +\hat{\rho}%
\left( \mathbf{r}_{2},t_{2}\right) \hat{\rho}\left( \mathbf{r}%
_{1},t_{1}\right) \right| vac\right\rangle .  \label{1.2a}
\end{equation}

If we integrate the Hamiltonian density operator, $\hat{\rho}\left( \mathbf{%
r,}t\right) ,$ with respect to $\mathbf{r}$ we get the Hamiltonian operator 
\begin{equation}
\hat{H}=\int_{V}\hat{\rho}\left( \mathbf{r,}t\right) d^{3}r.  \label{1.3}
\end{equation}
I will consider the volume $V$ as finite (e. g. a cube of side $V^{1/3})$,
but take the limit $V$ becoming the whole space at the end of the
calculations. I shall assume that the vacuum state is the eigenstate of $%
\hat{H}$ with the minimal eigenvalue, which I will label $E_{vac}$. That is 
\[
\hat{H}\mid vac\rangle =E_{vac}\mid vac\rangle ,E_{vac}=V\rho _{vac}, 
\]
with no other eigenvalue of $\hat{H}$ smaller than $E_{vac}$. Hence the $%
\mathbf{r}_{2}$ integral of the two-point correlation gives 
\begin{eqnarray}
\int_{V}C\left( \mathbf{r}_{1},t_{1};\mathbf{r}_{2},t_{2}\right) d^{3}r_{2}
&=&\left\langle vac\left| \hat{\rho}\left( \mathbf{r}_{1},t_{1}\right)
\int_{V}\hat{\rho}\left( \mathbf{r}_{2},t_{2}\right) d^{3}r_{2}\right|
vac\right\rangle  \nonumber \\
&=&\left\langle vac\left| \hat{\rho}\left( \mathbf{r}_{1}\mathbf{,}%
t_{1}\right) \hat{H}\right| vac\right\rangle  \nonumber \\
&=&\left\langle vac\left| \hat{\rho}\left( \mathbf{r}_{1}\mathbf{,}%
t_{1}\right) E_{vac}\right| vac\right\rangle =V\rho _{vac}^{2},  \label{1.4}
\end{eqnarray}
where eqs.$\left( \ref{1.2}\right) $ and $\left( \ref{1.3}\right) $ have
been taken into account. This equation contradicts eq.$\left( \ref{0.5}%
\right) ,$ the reason being that here we have used a different deffinition,
not making the assumption $\rho _{vac}=0.$ Actually the relevant quantity is
the two-point \textit{density correlation relative to the mean}, therefore I
will redefine the correlation after subtracting the square of the mean
density, that is 
\begin{equation}
C\left( \mathbf{r}_{1},t_{1};\mathbf{r}_{2},t_{2}\right) =C_{old}\left( 
\mathbf{r}_{1},t_{1};\mathbf{r}_{2},t_{2}\right) -\rho _{vac}^{2}.
\label{1.7}
\end{equation}
where $C_{old}$ is given by eq.$\left( \ref{1.2a}\right) .$ This correlation
does fulfil eq.$\left( \ref{0.5}\right) $.

A calculation of the energy density and the two-point correlation of all
quantum fields existing in nature (possibly those of the standard model of
high energy physics) is a formidable task, out of the scope ot this paper.
Here the study will be restricted to quantum electrodynamics (QED) in order
to get clues for the solution of the problems associated with the gravity of
the vacuum fields. That is I will consider a fictitious world where the
existing quantum fields are the electromagnetic one and the Dirac field of
electrons and positrons but no other fields.

\subsection{The free electromagnetic field. A paradox}

We will work in Minkowski space, a good enough approximation for the
calculation of the mean vacuum energy density. (But the approximation is not
valid when we take the two-point correlation into account, as will be
discussed in section 3.2). In the present subsection I will calculate the
stress-energy tensor of the free electromagnetic field of the vacuum,
ignoring all other vacuum fields. I shall assume homogeneity and isotropy,
so that the stress-energy tensor is defined by just two parameters, density $%
\rho $ and pressure $P$. However, as we shall see, it is not possible to
assume Lorentz invariance, that is $P=-\rho .$

The quantization of a classical field involves promoting the amplitudes of
the plane waves expansion to become (creation or annihilation) operators.
For instance the classical free electromagnetic field, in the Coulomb gauge,
may be represented by an expansion in plane waves of the vector potential, $%
\mathbf{A}(\mathbf{r,}t),$\ that is 
\begin{eqnarray}
\mathbf{A}(\mathbf{r,}t) &=&\frac{1}{\sqrt{V}}\sum_{\mathbf{k,\varepsilon }%
}\left( \frac{%TCIMACRO{\UNICODE[m]{0x127}}
%BeginExpansion
\rlap{\protect\rule[1.1ex]{.325em}{.1ex}}h%
%EndExpansion
}{2k}\right) ^{1/2}\times  \nonumber \\
&&\times \left[ \alpha _{\mathbf{k},\mathbf{\varepsilon }}\mathbf{%
\varepsilon }\exp \left( i\mathbf{k.r-}ikt\right) +\alpha _{\mathbf{k},%
\mathbf{\varepsilon }}^{*}\mathbf{\varepsilon }\exp \left( -i\mathbf{k.r+}%
ikt\right) \right] ,\smallskip  \label{2.1}
\end{eqnarray}
where $k=\left| \mathbf{k}\right| .$ From now on I will use the notation of
the book of Sakurai\cite{Sakurai} (except otherwise stated) and natural
units $%TCIMACRO{\UNICODE[m]{0x127}}
%BeginExpansion
\rlap{\protect\rule[1.1ex]{.325em}{.1ex}}h%
%EndExpansion
=c=1.$ However sometimes I will write explicitly Planck\'{}s constant $
%TCIMACRO{\UNICODE[m]{0x127}}
%BeginExpansion
\rlap{\protect\rule[1.1ex]{.325em}{.1ex}}h%
%EndExpansion
$ for clarity. From eq.$\left( \ref{2.1}\right) $ it is easy to get the
electric field, $\mathbf{E=-}\partial \mathbf{A/}\partial t$, and the
magnetic field, $\mathbf{B=\bigtriangledown \times A}$. The polarization
vector $\mathbf{\varepsilon }$ depends on $\mathbf{k}$ (in fact $\mathbf{%
k\cdot \varepsilon =}0)$ and it may have two possible values so that we
should write $\mathbf{\varepsilon }_{j}\left( \mathbf{k}\right) ,j=1,2,$ but
I will use a simplified notation except when some confussion might arise.

In the quantized field an annihilation operator $\hat{\alpha}_{\mathbf{k},%
\mathbf{\varepsilon }}$ is substituted for the amplitude $\alpha _{\mathbf{k}%
,\mathbf{\varepsilon }}$ and a creation operator $\hat{\alpha}_{\mathbf{k},%
\mathbf{\varepsilon }}^{\dagger },$ for the amplitude $\alpha _{\mathbf{k},%
\mathbf{\varepsilon }}^{*},$ whence the electric and magnetic fields, $%
\mathbf{\hat{E}}$ and $\mathbf{\hat{B},}$ become vector operators. They may
be written as an expansion in plane waves taking a quantized counterpart of
eq.$\left( \ref{2.1}\right) $ into account. From these expansions, that I do
not write explicitly, it is trivial to obtain the energy density operator.
We get 
\begin{eqnarray*}
\hat{\rho}_{EM} &\equiv &\frac{1}{2}\left( \mathbf{\hat{E}}^{2}+\mathbf{\hat{%
B}}^{2}\right) =\frac{1}{4V}\sum_{\mathbf{k,\varepsilon }}\sum_{\mathbf{k}%
^{\prime }\mathbf{,\varepsilon }^{\prime }}K_{1}(\hat{\alpha}_{\mathbf{k}%
^{\prime },\mathbf{\varepsilon }^{\prime }}^{\dagger }\hat{\alpha}_{\mathbf{k%
},\mathbf{\varepsilon }}+\hat{\alpha}_{\mathbf{k},\mathbf{\varepsilon }}\hat{%
\alpha}_{\mathbf{k}^{\prime },\mathbf{\varepsilon }^{\prime }}^{\dagger }) \\
&&+\frac{1}{4V}\sum_{\mathbf{k,\varepsilon }}\sum_{\mathbf{k}^{\prime }%
\mathbf{,\varepsilon }^{\prime }}\left( K_{2}\hat{\alpha}_{\mathbf{k},%
\mathbf{\varepsilon }}\hat{\alpha}_{\mathbf{k}^{\prime },\mathbf{\varepsilon 
}^{\prime }}+K_{2}^{*}\hat{\alpha}_{\mathbf{k},\mathbf{\varepsilon }%
}^{\dagger }\hat{\alpha}_{\mathbf{k}^{\prime },\mathbf{\varepsilon }^{\prime
}}^{\dagger }\right) ,
\end{eqnarray*}
where the functions $K_{1}$ and $K_{2}$ are c-numbers (not operators) given
by

\begin{equation}
K_{1}=\sqrt{kk^{\prime }}\mathbf{\varepsilon }*\mathbf{\varepsilon }^{\prime
}\exp \left[ i\left( \mathbf{k}-\mathbf{k}^{\prime }\right) \mathbf{.r-}%
i\left( k-k^{\prime }\right) t\right]  \label{2.2a}
\end{equation}
\begin{equation}
K_{2}=\sqrt{kk^{\prime }}\mathbf{\varepsilon }*\mathbf{\varepsilon }^{\prime
}\exp \left[ i\left( \mathbf{k}+\mathbf{k}^{\prime }\right) \mathbf{.r-}%
i\left( k+k^{\prime }\right) t\right] ,  \label{2.2b}
\end{equation}
For notational simplicity I have introduced the following ``star product'' 
\[
\mathbf{\varepsilon }*\mathbf{\varepsilon }^{\prime }\equiv \mathbf{%
\varepsilon }\cdot \mathbf{\varepsilon }^{\prime }+\frac{1}{kk^{\prime }}%
\left[ \left( \mathbf{k\times \varepsilon }\right) \mathbf{\cdot }\left( 
\mathbf{k}^{\prime }\mathbf{\times \varepsilon }^{\prime }\right) \right] , 
\]
with the properties 
\begin{eqnarray}
\sum_{j}(\mathbf{\varepsilon }_{i}*\mathbf{\varepsilon }_{j}^{\prime })(%
\mathbf{\varepsilon }_{j}^{\prime }*\mathbf{\varepsilon }_{l}^{\prime \prime
}) &=&\mathbf{\varepsilon }_{i}*\mathbf{\varepsilon }_{l}^{\prime \prime
},\sum_{ij}\mathbf{\varepsilon }_{i}*\mathbf{\varepsilon }_{j}=4  \nonumber
\\
\sum_{ij}(\mathbf{\varepsilon }_{i}*\mathbf{\varepsilon }_{j}^{\prime })^{2}
&=&2\left[ 1+\frac{\mathbf{k\cdot k}^{\prime }}{kk^{\prime }}\right] ^{2}.
\label{2.2d}
\end{eqnarray}

It is convenient to write all terms of $\hat{\rho}_{EM}$ in normal order,
that is the annihilation (creation) operators to the right (left). Taking
the commutation rules of the field operators into account we obtain 
\begin{eqnarray}
\hat{\rho}_{EM} &=&\rho _{EM0}+\hat{\rho}_{EM1}+\hat{\rho}_{EM2},
\label{2.2c} \\
\hat{\rho}_{EM1} &=&\frac{1}{2V}\sum_{\mathbf{k,\varepsilon }}\sum_{\mathbf{k%
}^{\prime }\mathbf{,\varepsilon }^{\prime }}K_{1}\hat{\alpha}_{\mathbf{k}%
^{\prime },\mathbf{\varepsilon }^{\prime }}^{\dagger }\hat{\alpha}_{\mathbf{k%
},\mathbf{\varepsilon }},  \nonumber \\
\hat{\rho}_{EM2} &=&\frac{1}{4V}\sum_{\mathbf{k,\varepsilon }}\sum_{\mathbf{k%
}^{\prime }\mathbf{,\varepsilon }^{\prime }}\left( K_{2}\hat{\alpha}_{%
\mathbf{k},\mathbf{\varepsilon }}\hat{\alpha}_{\mathbf{k}^{\prime },\mathbf{%
\varepsilon }^{\prime }}+K_{2}^{*}\hat{\alpha}_{\mathbf{k},\mathbf{%
\varepsilon }}^{\dagger }\hat{\alpha}_{\mathbf{k}^{\prime },\mathbf{%
\varepsilon }^{\prime }}^{\dagger }\right) ,  \nonumber
\end{eqnarray}
where $\rho _{EM0}$ is a numerical constant (times de unit operator) given
by 
\[
\rho _{EM0}=\frac{1}{2V}\sum_{\mathbf{k,\varepsilon }}k. 
\]

The Hamiltonian is obtained by performing a space integral of the energy
density eq.$\left( \ref{2.2c}\right) $, that is 
\begin{eqnarray}
\hat{H}_{EM} &=&\lim_{V\rightarrow \infty }\int_{V}\hat{\rho}_{EM}\left( 
\mathbf{r}\right) d^{3}\mathbf{r}  \nonumber \\
&=&\sum_{\mathbf{k\varepsilon }}k(\hat{\alpha}_{\mathbf{k},\mathbf{%
\varepsilon }}^{\dagger }\hat{\alpha}_{\mathbf{k},\mathbf{\varepsilon }}+%
\frac{1}{2}).  \label{HEM}
\end{eqnarray}
We see that the integral cancels de $V$ denominator and removes the
spacetime dependence.

For the free electromagnetic field the vacuum state, $\mid 0\rangle ,$ may
be defined as the state with the minimal energy, that is the smallest
eigenvector of the operator eq.$\left( \ref{HEM}\right) .$ It is a state
with zero photons and it has the properties 
\[
\alpha _{\mathbf{k},\mathbf{\varepsilon }}\mid 0\rangle =0,\langle 0\mid
\alpha _{\mathbf{k},\mathbf{\varepsilon }}^{\dagger }=0. 
\]
The state $\mid 0\rangle $ is different from the actual QED vacuum, $\mid
vac\rangle ,$ that takes the interaction with the electron-positron field
into account. It will be studied below. For the purely electromagnetic
vacuum state $\mid 0\rangle $ the expectation of the energy density is 
\begin{equation}
\left\langle 0\left| \hat{\rho}_{EM}\right| 0\right\rangle =\rho _{EM0}=%
\frac{1}{V}\sum_{\mathbf{k,\varepsilon }}\frac{1}{2} 
%TCIMACRO{\UNICODE[m]{0x127}}
%BeginExpansion
\rlap{\protect\rule[1.1ex]{.325em}{.1ex}}h%
%EndExpansion
k=\frac{1}{V}\sum_{\mathbf{k}}%TCIMACRO{\UNICODE[m]{0x127}}
%BeginExpansion
\rlap{\protect\rule[1.1ex]{.325em}{.1ex}}h%
%EndExpansion
k,  \label{2.4}
\end{equation}
where the latter equality derives from the two possible polarizations. We
have taken into account that the two latter terms of eq.$\left( \ref{2.2c}%
\right) $ do not contribute to the expectation value because either the
annihilation operator placed on the right or the creation operator on the
left gives zero when acting on the vacuum state. In the limit $V\rightarrow
\infty $ we obtain

\begin{eqnarray}
\rho _{EM0} &=&\frac{1}{V}\sum_{\mathbf{k}}k\rightarrow \int k\left( 2\pi
\right) ^{-3}d^{3}k,  \nonumber \\
&=&\frac{1}{2\pi ^{2}}\int_{0}^{k_{\max }}k^{3}dk=\frac{k_{\max }^{4}}{8\pi
^{2}},  \label{2.5}
\end{eqnarray}
where we have introduced a cut-off, $k_{\max }^{4},$ in the wavevectors (or
the momenta, what is equivalent in our units). We see that the integral is
strongly divergent in the limit $k_{\max }\rightarrow \infty ,$ but it is
plausible that a cut-off could be originated by fluctuations of the
spacetime metric, as will be discussed in section 3.2.

In order to define the stress-energy tensor we need the pressure. As is well
known the pressure is $1/3$ times the energy density for any isotropic
electromagnetic radiation. Therefore the pressure of the vacuum
electromagnetic field would be one third the quantity eq.$\left( \ref{2.5}%
\right) .$ Hence the stress-energy tensor of the free electromagnetic field
of the vacuum is diagonal, and its nonzero components are 
\begin{equation}
\rho _{EM0}=\frac{k_{\max }^{4}}{8\pi ^{2}},P_{EM0}=\frac{k_{\max }^{4}}{%
24\pi ^{2}},\smallskip  \label{TEM}
\end{equation}
That is the tensor is traceless, fulfilling 
\[
T_{\mu }^{\mu }=3P_{EM0}-\rho _{EM0}=0, 
\]
for any choice of the cut-off wavevector $k_{\max }.$

This gives rise to a paradox because it is plausible that the stress-energy
tensor of the vacuum is proportional to the metric tensor, as a consequence
of the Lorentz invariance that we should assume for the vacuum (in Minkowski
space). Equivalently the tensor in mixed coordinates should be diagonal with
a pressure equal to \emph{minus} the energy density. Therefore if the vacuum
fields are real, which is the fundamental hypothesis of this paper, there is
a contradiction between the stress-energy tensor predicted by the Maxwell
equations (for any free radiation field) and the Lorentz invariance of the
vacuum fields. Actually it may be argued that neither the density nor the
pressure are well defined, eqs.$\left( \ref{TEM}\right) $ being divergent
(or cut-off dependent), but there is a problem in any case. I think that the
plausible solution is that, as the vacuum contains many fields, \textit{the
stress-energy tensor of all interacting fields toghether is well defined and
Lorentz invariant}, but the tensor of every free field alone is not. (The
divergences might be removed by an effective cut-off, see section 3.2 ).
Studying the stress-energy tensor of all fields is beyond the scope of this
paper, but a support to our hypothesis results from the study of the
electron-positron field to be made in the next subsection.

\subsection{The electron-positron vacuum field}

We need the stress-energy tensor of the Dirac field, that is the energy
density, $\rho ,$ and the pressure, $P$. In order to avoid any confussion
caused by the different notations used in the literature, I will write $\rho 
$ and $P$ in terms of the original Dirac matrices $\alpha _{k}$ and $\beta $
(rather than the gamma matrices that have been defined in several different
forms, compare e. g. \cite{Sakurai} and \cite{Schweber}). Thus the
corresponding operators may be written 
\begin{equation}
\hat{\rho}_{D}=\frac{i}{2}\left( \hat{\psi}^{\dagger }\frac{d\hat{\psi}}{dt}-%
\frac{d\hat{\psi}^{\dagger }}{dt}\hat{\psi}\right) ,  \label{2.5b}
\end{equation}
\begin{equation}
\hat{P}_{D}^{(k}=\frac{i}{2}\left( \hat{\psi}^{\dagger }\alpha _{j}\frac{%
\partial \hat{\psi}}{\partial x^{j}}-\frac{\partial \hat{\psi}^{\dagger }}{%
\partial x^{j}}\alpha _{j}\hat{\psi}\right) ,  \label{2.5a}
\end{equation}
where $\hat{\psi}$ and $\hat{\psi}^{\dagger }$ are quantized fields and $j$
may be either 1, 2 or 3, the resulting pressure being the same in the three
cases due to the assumed isotropy of the vacuum. Expanding $\hat{\psi}$ and $%
\hat{\psi}^{\dagger }$ in plane waves we get 
\begin{eqnarray*}
\hat{\psi}\left( \mathbf{r},t\right) &=&\sqrt{\frac{1}{V}}\sum_{\mathbf{p},s}%
\sqrt{\frac{m}{E}}\hat{b}_{\mathbf{p}s}u_{\mathbf{p}s}\exp \left( i\mathbf{%
p\cdot r}-iEt\right) \\
&&+\sqrt{\frac{1}{V}}\sum_{\mathbf{p},s}\sqrt{\frac{m}{E}}\hat{d}_{\mathbf{p}%
s}^{\dagger }v_{\mathbf{p}s}\exp \left( -i\mathbf{p\cdot r}+iEt\right) ,
\end{eqnarray*}
\begin{eqnarray*}
\hat{\psi}^{\dagger }\left( \mathbf{r},t\right) &=&\sqrt{\frac{1}{V}}\sum_{%
\mathbf{p},s}\sqrt{\frac{m}{E}}\hat{b}_{\mathbf{p}s}^{\dagger }u_{\mathbf{p}%
s}^{\dagger }\exp \left( -i\mathbf{p\cdot r}+iEt\right) \\
&&+\sqrt{\frac{1}{V}}\sum_{\mathbf{p},s}\sqrt{\frac{m}{E}}\hat{d}_{\mathbf{p}%
s}v_{\mathbf{p}s}^{\dagger }\exp \left( i\mathbf{p\cdot r}-iEt\right) ,
\end{eqnarray*}
where $\hat{b}_{\mathbf{p},s}$ $\left( \hat{d}_{\mathbf{p},s}\right) $ is
the annihilation operator of an electron (positron) with momentum $\mathbf{p}
$ and spin $s(=1,2)$, and $\hat{b}_{\mathbf{p},s}^{\dagger }\left( \hat{d}_{%
\mathbf{p},s}^{\dagger }\right) $ the corresponding creation operator and $%
u,u^{\dagger },v,v^{\dagger }$ are appropriate spinors. (Our notation
follows the book of Sakurai\cite{Sakurai}.) Inserting these expressions in
eq.$\left( \ref{2.5b}\right) $ we get $\hat{\rho}_{D}$ as a sum of 4 terms
with a product of 2 operators each. We want all the terms with the creation
and annihilation operators in normal order so that, taking the
anticommutation rules into account, we should make a replacement as follows 
\begin{eqnarray*}
&&\hat{d}_{\mathbf{p}s}\exp \left( i\mathbf{p\cdot r}-iEt\right) \hat{d}_{%
\mathbf{p}^{\prime }s^{\prime }}^{\dagger }\exp \left( -i\mathbf{p}^{\prime }%
\mathbf{\cdot r}+iE^{\prime }t\right) \\
&=&\delta _{\mathbf{pp}^{\prime }}\delta _{ss^{\prime }}-\hat{d}_{\mathbf{p}%
^{\prime }s^{\prime }}^{\dagger }\exp \left( -i\mathbf{p}^{\prime }\mathbf{%
\cdot r}+iE^{\prime }t\right) \hat{d}_{\mathbf{p}s}\exp \left( i\mathbf{%
p\cdot r}-iEt\right) ,
\end{eqnarray*}
where $\delta _{\mathbf{pp}^{\prime }}$ and $\delta _{ss^{\prime }}$ are
Kronecker's deltas. Thus we obtain 5 terms, all in normal order, that is 
\begin{equation}
\hat{\rho}_{D}\left( \mathbf{r},t\right) =\rho _{D0}+\hat{\rho}_{b}\left( 
\mathbf{r},t\right) +\hat{\rho}_{d}\left( \mathbf{r},t\right) +\hat{\rho}%
_{bd}\left( \mathbf{r},t\right) +\hat{\rho}_{bd}^{\dagger }\left( \mathbf{r}%
,t\right) .  \label{2.6}
\end{equation}
The latter 4 terms of eq.$\left( \ref{2.6}\right) $ are as follows 
\begin{eqnarray*}
\hat{\rho}_{b}\left( \mathbf{r},t\right) &=&\frac{1}{V}\sum_{\mathbf{pp}%
^{\prime }ss^{\prime }}B\left( \mathbf{p,}s,\mathbf{p}^{\prime },s^{\prime
}\right) \hat{b}_{\mathbf{p}^{\prime }s^{\prime }}^{\dagger }b_{\mathbf{p}%
s}\exp \left[ i\left( \mathbf{p-p}^{\prime }\right) \cdot \mathbf{r-}i\left(
E-E^{\prime }\right) t\right] , \\
\hat{\rho}_{d}\left( \mathbf{r},t\right) &=&\frac{1}{V}\sum_{\mathbf{pp}%
^{\prime }ss^{\prime }}D\left( \mathbf{p,}s,\mathbf{p}^{\prime },s^{\prime
}\right) d_{\mathbf{p}^{\prime }s^{\prime }}^{\dagger }d_{\mathbf{p}s}\exp
\left[ i\left( \mathbf{p-p}^{\prime }\right) \cdot \mathbf{r-}i\left(
E-E^{\prime }\right) t\right] \\
\hat{\rho}_{bd}\left( \mathbf{r},t\right) &=&\frac{1}{V}\sum_{\mathbf{pp}%
^{\prime }ss^{\prime }}F\left( \mathbf{p,}s,\mathbf{p}^{\prime },s^{\prime
}\right) d_{\mathbf{p}s}b_{\mathbf{p}^{\prime }s^{\prime }}\exp \left[
i\left( \mathbf{p+p}^{\prime }\right) \cdot \mathbf{r-}i\left( E+E^{\prime
}\right) t\right] , \\
\hat{\rho}_{bd}^{\dagger }\left( \mathbf{r},t\right) &=&\frac{1}{V}\sum_{%
\mathbf{pp}^{\prime }ss^{\prime }}F^{*}\left( \mathbf{p,}s,\mathbf{p}%
^{\prime },s^{\prime }\right) b_{\mathbf{p}^{\prime }s^{\prime }}^{\dagger
}d_{\mathbf{p}s}^{\dagger }\exp \left[ -i\left( \mathbf{p+p}^{\prime
}\right) \cdot \mathbf{r+}i\left( E+E^{\prime }\right) t\right] .
\end{eqnarray*}
Getting the functions $B,D,F$ and $F^{*}$ is straightforward but we report
only the expression of $\left| F\right| ^{2}$ after the spin sum, which will
be used in section 3. We get 
\begin{eqnarray}
\sum_{ss^{\prime }}\left| F\right| ^{2} &=&\frac{m^{2}}{4EE^{\prime }}%
(E-E^{\prime })^{2}\sum_{ss^{\prime }}\left| v_{s}^{\dagger }\left( \mathbf{p%
}\right) u_{s^{\prime }}\left( \mathbf{p}^{\prime }\right) \right| ^{2} 
\nonumber \\
&=&\frac{m^{2}}{4}\left( E-E^{\prime }\right) ^{2}\left[ 1+\frac{\mathbf{%
p\cdot p}^{\prime }-m^{2}}{EE^{\prime }}\right] ,  \label{F2}
\end{eqnarray}
where we have taken into account the expressions of the spinors $u$ and $v$%
\cite{Sakurai}. This leads to

\begin{equation}
\sum_{ss^{\prime }}\left| v_{s}^{\dagger }\left( \mathbf{p}\right)
u_{s^{\prime }}\left( \mathbf{q}\right) \right| ^{2}=\frac{\left( E+m\right)
\left( E^{\prime }+m\right) }{2m^{2}}\left| \frac{\mathbf{p}}{E+m}+\frac{%
\mathbf{q}}{E^{\prime }+m}\right| ^{2},  \label{9}
\end{equation}
whence putting 
\begin{eqnarray*}
\mathbf{p}^{2} &=&E^{2}-m^{2}=\left( E+m\right) \left( E-m\right) , \\
\mathbf{q}^{2} &=&E^{\prime 2}-m^{2}=\left( E^{\prime }+m\right) \left(
E^{\prime }-m\right) ,
\end{eqnarray*}
it is easy to get eq.$\left( \ref{F2}\right) .$

The first term in eq.$\left( \ref{2.6}\right) $, $\rho _{D0},$ is a
c-number, not an operator (more properly it is proportional to the unit
operator), that is 
\begin{equation}
\rho _{D0}=-\frac{1}{V}\sum_{p\mathbf{,}s}\sqrt{m^{2}+p^{2}}.  \label{roD0}
\end{equation}

Integration of eq.$\left( \ref{2.6}\right) $ with respect to $\mathbf{r}$%
\textbf{\ }gives the Hamiltonian of the free electron-positron field, that is

\begin{equation}
H_{D}=\sum_{p\mathbf{,}s}\sqrt{m^{2}+p^{2}}\left( b_{\mathbf{p},s}^{\dagger
}b_{\mathbf{p},s}+d_{\mathbf{p},s}^{\dagger }d_{\mathbf{p},s}-1\right) .
\label{HD}
\end{equation}
The vacuum state, $\mid 0\rangle ,$ of the free field corresponds to the
eigenvector of the Hamiltonian eq.$\left( \ref{HD}\right) $ with the
smallest eigenvalue. It consists of zero electrons and zero positrons.
Therefore from now on we will define $\mid 0\rangle $ to be the QED
unperturbed vacuum state, which is a simultaneous eigenvector of both
Hamiltonians eqs.$\left( \ref{HEM}\right) $ and $\left( \ref{HD}\right) .$
Thus the state $\mid 0\rangle $ is defined, from now on, as having zero
photons, electrons and positrons. It should be distinguished from the
physical vacuum state, $\mid vac\rangle ,$ which is an eigenvalue of the
total Hamiltonian, including the interactions.

Our aim is to obtain the energy density, that is the expectation value in
the vacuum of the density operator eq.$\left( \ref{2.6}\right) .$ Neglecting
the field interactions (to be studied in the next subsection), we get the
energy density of the unperturbed electron-positron field as the vacuum
expectation of the operator, that is 
\[
\left\langle 0\left| \hat{\rho}_{D}\left( \mathbf{r},t\right) \right|
0\right\rangle =\left\langle 0\left| \hat{\rho}_{D0}\left( \mathbf{r}%
,t\right) \right| 0\right\rangle =\rho _{D0} 
\]
The former equality follows from the fact that the four latter terms of eq.$%
\left( \ref{2.6}\right) $ do not contribute to the expectation because
either the annihilation operator placed on the right or the creation
operator on the left gives zero when acting on the vacuum state.

The quantity $\rho _{D0}$, eq.$\left( \ref{roD0}\right) $, leads to the
following \textit{negative} divergent energy density in the limit $%
V\rightarrow \infty $%
\begin{eqnarray}
\rho _{D0} &=&-\left( 2\pi \right) ^{-3}\sum_{s}\int_{0}^{p_{\max
}}Ed^{3}p=-\pi ^{-2}\int_{0}^{p_{\max }}\sqrt{m^{2}+p^{2}}p^{2}dp  \nonumber
\\
&=&-\frac{1}{8\pi ^{2}}\left[ p_{\max }\sqrt{m^{2}+p_{\max }^{2}}\left(
m^{2}+2p_{\max }^{2}\right) -m^{4}\arg \sinh \left( \frac{p_{\max }}{m}%
\right) \right]  \nonumber \\
&=&-\frac{1}{4\pi ^{2}}\left[ p_{\max }^{4}+p_{\max }^{2}m^{2}+\frac{1}{8}%
m^{4}-\frac{1}{2}m^{4}\ln \left( \frac{2p_{\max }}{m}\right) \right]
+O\left( p_{\max }^{-2}\right) ,\smallskip  \label{2.11}
\end{eqnarray}
where we have introduced an ultraviolet cut-off of the momenta, $p_{\max }.$
I point out that the negative value might be anticipated by inspection of
the Hamiltonian eq.$\left( \ref{HD}\right) .$ In fact the density $\rho
_{D0} $ is just the (unperturbed) vacuum expectation of the Hamiltonian
divided by the volume $V$.

The stresses may be calculated as vacuum expectations of the operators $\hat{%
P}_{D}^{(k},$ eq.$\left( \ref{2.5a}\right) $. As in the case of the energy
density, eq.$\left( \ref{2.6}\right) ,$ we may write the operator $\hat{P}%
_{D}^{(k}$ as a sum of terms all of them with the creation and annihilation
operators in normal order. We get 
\begin{equation}
\hat{P}_{D}^{(k}\left( \mathbf{r},t\right) =P_{D0}^{(k}+\hat{P}%
_{b}^{(k}\left( \mathbf{r},t\right) +\hat{P}_{d}^{(k}\left( \mathbf{r}%
,t\right) +\hat{P}_{bd}^{(k}\left( \mathbf{r},t\right) +\hat{P}%
_{bd}^{(k\dagger }\left( \mathbf{r},t\right) .  \label{2.13}
\end{equation}
The latter 4 terms are similar to those of eq.$\left( \ref{2.6}\right) $ but
I will not give their explicit expressions. These 4 terms do not contribute
to the mean pressure. The former term of eq.$\left( \ref{2.13}\right) $ does
contribute and, after some algebra, it becomes 
\begin{eqnarray*}
P_{D0}^{(k} &=&\frac{m}{V}\sum_{\mathbf{p}s}\frac{p_{k}}{E}v_{\mathbf{p}%
s}^{\dagger }\alpha _{k}v_{\mathbf{p}s}=\frac{2}{V}\sum_{\mathbf{p}}\frac{%
p_{k}^{2}}{E} \\
&\rightarrow &\frac{1}{\pi ^{2}}\int_{0}^{p_{\max }}\frac{p_{k}^{2}}{E}%
p^{2}dp=\frac{1}{3\pi ^{2}}\int_{0}^{p_{\max }}\frac{p^{4}}{E}dp,
\end{eqnarray*}
where the isotropy has been taken into account in the latter equality. That
is the stresses along three orthogonal axes are equal and every one
corresponds to the pressure. We get 
\[
P_{D0}=\frac{1}{12\pi ^{2}}\left[ \left( p_{\max }^{3}-\frac{3}{2}%
m^{2}p_{\max }\right) \sqrt{m^{2}+p_{\max }^{2}}+\frac{3m^{4}}{2}\arg \sinh
\left( \frac{p_{\max }}{m}\right) \right] . 
\]

The sum, $\rho _{ZPF}$ (for `zeropoint field')$,$ of the quantities eqs.$%
\left( \ref{2.5}\right) $ and $\left( \ref{2.11}\right) $ is the energy
density of the QED unperturbed vacuum state. It is not obvious whether we
should identify the cut-off photon momentum (the same as the energy in
natural units), $k_{\max }$, with either the maximum electron momentum $%
p_{\max }$ or the maximum electron energy, $E_{\max }\equiv \sqrt{%
m^{2}+p_{\max }^{2}}.$ With both choices the leading term of $\rho _{ZPF}$
becomes 
\begin{equation}
\rho _{ZPF}\equiv \rho _{EM0}+\rho _{D0}=-\frac{k_{\max }^{4}}{8\pi ^{2}}%
+O\left( k_{\max }^{2}\right) ,  \label{ZPF}
\end{equation}
That energy density $\rho _{ZPF}$ is the quantity removed in standard
quantum calculations by the normal ordering rule. The sum of the pressures
is positive, namely 
\begin{equation}
P_{ZPF}\equiv P_{EM0}+P_{D0}=\frac{k_{\max }^{4}}{8\pi ^{2}}+O\left( k_{\max
}^{2}\right) .  \label{PZPF}
\end{equation}
More precisely, neglecting only terms that go to zero when the cut-offs go
to infinity, we get 
\begin{eqnarray}
P_{ZPF}+\rho _{ZPF} &=&\frac{1}{6\pi ^{2}}\left[ k_{\max }^{4}-p_{\max
}^{3}E_{\max }\right]  \nonumber \\
&&+\frac{m^{4}}{4\pi ^{2}}\arg \sinh \left( \frac{p_{\max }}{m}\right) ,
\label{perro}
\end{eqnarray}
that suggests the identification 
\[
k_{\max }^{4}=p_{\max }^{3}E_{\max }. 
\]

It is remarkable that the \textit{leading terms give rise to a stress-energy
tensor fulfilling the Lorentz invariant relation }$P_{ZPF}=-\rho _{ZPF}$,
although every vacuum field alone does not fulfil that relation. On the
other hand the sum eq.$\left( \ref{perro}\right) $ has \textit{positive
pressure and negative energy density }contrary to what we might expect. In
section 4 I will propose a solution for this anomaly. The relation eq.$%
\left( \ref{perro}\right) $ corresponds to the neglect of the interaction,
to be studied in the next subsection, and it has been calculated only for
the particular case of QED. Therefore no strong conclusion may be obtained,
but our result suggests that the stress-energy tensor of all interacting
fields, might be Lorentz invariant in the vacuum (in Minkowski space).

\subsection{Corrections for the interaction}

The physical vacuum of QED, $\mid vac\rangle ,$ is different from the zeroth
order vacuum, $\mid 0\rangle ,$ studied above. The latter is the
eigenvector, with the smallest eigenvalue, of the unperturbed Hamiltonian $%
H_{0}=H_{EM}+H_{D},$ see eqs.$\left( \ref{HEM}\right) $ and $\left( \ref{HD}%
\right) .$ The former is the eigenvector of $H=H_{EM}+H_{D}+H_{int}$ with
the smallest eigenvalue, that is taking the interactions into account.
Finding $\mid vac\rangle $ as an exact eigenvector of $H$ is not possible
and we should use a perturbation method, that is to calculate it as an
expansion in powers of the coupling constant, the positron charge, $e$.
Actually only even powers of $e$ would appear and the result becomes an
expansion in powers of the fine structure constant $\alpha \equiv
e^{2}/(4\pi %TCIMACRO{\UNICODE[m]{0x127}}
%BeginExpansion
\rlap{\protect\rule[1.1ex]{.325em}{.1ex}}h%
%EndExpansion
c)\simeq 1/137.$

The interaction Hamiltonian (or energy) operator may be written, in the
Coulomb gauge, 
\begin{equation}
\hat{\rho}_{int}\left( \mathbf{r,}t\right) =-e\hat{\psi}^{\dagger }\mathbf{%
\alpha }\hat{\psi}\cdot \mathbf{\hat{A}.}  \label{ront}
\end{equation}
The operators $\hat{\psi},\hat{\psi}^{\dagger }$ and $\mathbf{\hat{A}}$
contain two terms each when expanded in plane waves, each term corresponding
to an infinite sum. One of these terms has creation operators and the other
one annihilation operators (see e. g. eq.$\left( \ref{2.1}\right) $ for the
electromagnetic potential vector). This gives rise to 8 terms for $\hat{\rho}%
_{int},$ eq.$\left( \ref{ront}\right) $. Integrating with respect to $%
\mathbf{r}$ inside the volume $V$ leads to the interaction Hamiltonian. Only
two terms survive and we get 
\begin{eqnarray}
\hat{H}_{int} &=&-e\sum_{\mathbf{p,q},\mathbf{k},s,s^{\prime },\varepsilon }%
\frac{m}{V^{3/2}\sqrt{2kEE^{\prime }}}u_{s}^{\dagger }\left( \mathbf{p}%
\right) \mathbf{\alpha \cdot \varepsilon }v_{s^{\prime }}\left( \mathbf{q}%
\right)  \nonumber \\
&&\times \delta _{\mathbf{p+q,k}}\hat{\alpha}_{\mathbf{k},\mathbf{%
\varepsilon }}^{\dagger }b_{\mathbf{p},s}^{\dagger }d_{\mathbf{q},s^{\prime
}}^{\dagger }+h.c,\mathbf{.}  \label{Hint}
\end{eqnarray}
where $\delta _{\mathbf{p+q,k}}$ is a Kronecker's delta, $h.c.$ means
Hermitean conjugate and 
\[
E\equiv \sqrt{p^{2}+m^{2}},E^{\prime }\equiv \sqrt{q^{2}+m^{2}}. 
\]
One of the terms of the Hamiltonian may create triples
electron-positron-photon and the other term may annihilate triples. The
integral in $\mathbf{r}$ causes that the Hamiltonian eq.$\left( \ref{Hint}%
\right) $ is invariant under traslations and rotations, whence it can couple
only states with the same total momentum and total angular momentum as shown
below, see eqs.$\left( \ref{2.17}\right) $ and $\left( \ref{cj}\right) .$

The physical vacuum, $\mid vac\rangle ,$ may be obtained as an expansion in
powers of the coupling constant $\alpha $, but for the calculation of the
energy to first order we need $\mid vac\rangle $ only to order $\alpha
^{1/2},$ that is we should not consider states with more than three
particles (i. e. one electron, one positron and one photon). We have to that
order 
\begin{equation}
\mid vac\rangle =N\left[ \mid 0\rangle +\sum_{j}c_{j}\mid j\rangle \right] ,
\label{2.17}
\end{equation}
where $\mid j\rangle $ represents a state with one photon-electron-positron
triple (different $j$ correspond to the particles having either different
momenta or spins or both). $N$ is a normalization factor. The coefficients $%
c_{j}$ may be got from the interaction Hamiltonian, that is 
\begin{equation}
c_{j}=\frac{\left\langle j\left| \hat{H}_{int}\right| 0\right\rangle }{%
E_{0}-E_{j}},E_{j}=\left\langle j\left| \hat{H}_{0}\right| j\right\rangle
=V\rho _{j}.  \label{cj}
\end{equation}
By the properties of $H_{int}$, see above, the states $\mid j\rangle $ have
the same net momentum and angular momentum than the vacuum, that is zero.
However they possess positive energy above the vacuum state $\mid 0\rangle .$

The three-particles state may be got from $\mid 0\rangle $ by the action of
creation operators as follows 
\begin{equation}
\mid j\rangle =\hat{a}_{\mathbf{k}\varepsilon }^{\dagger }\hat{b}_{\mathbf{p}%
s}^{\dagger }\hat{d}_{\mathbf{q}s^{\prime }}^{\dagger }\mid 0\rangle
,j\equiv \left\{ \mathbf{k,}\varepsilon ,\mathbf{p,}s,\mathbf{q,}s^{\prime
}\right\} ,  \label{j}
\end{equation}
with the constraints 
\[
\mathbf{k+p+q}=0,\mathbf{\varepsilon +s+s}^{\prime }=0. 
\]
The unperturbed energy of one of these states is the corresponding
eigenvalue of the unperturbed Hamiltonian, that is 
\begin{equation}
E_{j}=k+\sqrt{m^{2}+\mathbf{p}^{2}}+\sqrt{m^{2}+\mathbf{q}^{2}}+E_{ZPF},
\label{Ej}
\end{equation}
where $E_{ZPF}$ is the vacuum energy in the volume $V$, that is the density $%
\rho _{ZPF},$ eq.$\left( \ref{ZPF}\right) ,$ times the volume $V$.
Corrections of higher order in $\alpha $ would involve a sum similar to eq.$%
\left( \ref{2.17}\right) ,$ but containing more general states, that is
states having $n$ triples electron-positron-photon, $n=1,2,3,...$All those
states, in addition to having nil momentum and angular momentum, possess the
quantum numbers of the unperturbed vacuum state, $\mid 0\rangle ,$ i. e.
zero electric charge (and zero leptonic number). Thus the vacuum state $\mid
vac\rangle $ also possesses those properties.

There are other states defined by a sum similar to eq.$\left( \ref{2.17}%
\right) ,$ but orthogonal to the vacuum state $\mid vac\rangle .$ That is
states of the form 
\[
\mid vac^{(l}\rangle =\sum_{j}c_{j}^{(l}\mid j\rangle , 
\]
where now $\mid j\rangle $ represents a general state with $n$ triples
electron-positron-photon, $j=0$ corresponding to the unperturbed vacuum
state $\mid 0\rangle .$ Thus for clarity I will rewrite the vacuum state eq.$%
\left( \ref{2.17}\right) $ in the form (not normalized) 
\[
\mid vac\rangle =\sum_{j}c_{j}\mid j\rangle . 
\]
Amongst the states $\mid vac^{(l}\rangle $ orthogonal to $\mid vac\rangle ,$
that is fulfilling 
\[
\sum_{j}c_{j}^{*}c_{j}^{(l}=0, 
\]
the one with the minimal energy will be named ``first excited vacuum
state'', labelled with $l=1$. Similarly the ``second excited vacuum state'', 
$l=2$, will be the state with minimal energy amongst those orthogonal to
both the vacuum state and the first excited vacuum state, and so on.

It is common practice in quantum field theory to name ``excitations of the
vacuum'' the states with particles. For instance a state with one electron
or a state with one electron and one photon, etc. However the states here
called ``excited vacuum states'' are a particular kind of vacuum
excitations, namely those characterized by having the same quantum numbers
of the vacuum (e. g. zero charge if we restrict ourselves to QED) and also
zero net momentum and net angular momentum. An interesting question is
whether these states appear somewhere in nature. My conjecture is that, with
some slight modification, they may be the constituents of the ``dark
matter'', as will be commented in section 5.

\section{Two-point correlations of the vacuum fields}

Vacuum fluctuations may be characterized by the two-point correlation of the
vacuum stress-energy tensor. Assuming homogeneity and isotropy that tensor
contains at most two independen parameters, namely energy density and
pressure. However the correlations of the different components of the
stress-energy tensor may depend on the direction. E. g. the correlation $%
\left\langle T_{xx}(\mathbf{r}_{a})T_{yy}\left( \mathbf{r}_{b}\right)
\right\rangle $ may depend on the angles of the vector $\mathbf{r}_{a}-%
\mathbf{r}_{b}$ with the directions $X$ and $Y$. In this paper I will not
calculate the two-point correlations of the vacuum fields in detail. But for
illustrative purposes I will get the correlation of the density at equal
times, which depends on a single parameter, namely $\left| \mathbf{r}_{a}-%
\mathbf{r}_{b}\right| $. I shall do that in the approximation of zeroth
order, that is involving quantum expectations in the unperturbed vacuum
state $\mid 0\rangle ,$ rather than the physical vacuum state $\mid
vac\rangle .$

\subsection{Two-point density correlation for free vacuum fields}

In this subsection I will work in Minkowski space, but see next subsection
for a discussion about the validity of this assumption. Using Cartesian
coordinates I shall calculate the correlation eq.$\left( \ref{1.7}\right) ,$
that is 
\begin{equation}
C(r)\equiv \left\langle vac\left| \hat{\rho}\left( \mathbf{r}_{1}\right) 
\hat{\rho}\left( \mathbf{r}_{2}\right) \right| vac\right\rangle -\rho
^{2},\rho \equiv \left\langle vac\left| \hat{\rho}\right| vac\right\rangle .
\label{3.11}
\end{equation}
The energy density operator is the sum of 3 terms, namely 
\begin{equation}
\hat{\rho}=\hat{\rho}_{EM}+\hat{\rho}_{D}+\hat{\rho}_{int},  \label{3.12}
\end{equation}
that have been discussed in section 2, see eqs.$\left( \ref{2.2c}\right) $
and $\left( \ref{2.6}\right) $ and $\left( \ref{ront}\right) .$

I start proving that a numerical constant added to the energy density
operator does not change the value of $C(r)$. In fact, if we substitute $%
\hat{\rho}+K$ for $\hat{\rho}$ in eq.$\left( \text{\ref{3.11}}\right) $ we
get 
\begin{eqnarray*}
C^{\prime }(r) &\equiv &\left\langle vac\left| \left[ \hat{\rho}\left( 
\mathbf{r}_{1}\right) +K\right] \left[ \hat{\rho}\left( \mathbf{r}%
_{2}\right) +K\right] \right| vac\right\rangle -\left\langle vac\left|
\left[ \hat{\rho}+K\right] \right| vac\right\rangle ^{2} \\
&=&C(r)+K\left[ \left\langle vac\left| \hat{\rho}\left( \mathbf{r}%
_{1}\right) \right| vac\right\rangle +\left\langle vac\left| \hat{\rho}%
\left( \mathbf{r}_{2}\right) \right| vac\right\rangle \right]
-2K\left\langle vac\left| \hat{\rho}\right| vac\right\rangle \\
&=&C(r),
\end{eqnarray*}
where we have taken into account that $\left\langle vac\mid vac\right\rangle
=1$ and $\left\langle vac\left| \hat{\rho}\left( \mathbf{r}_{1}\right)
\right| vac\right\rangle $ cannot depend on $\mathbf{r}$ due to the
traslational invariance of the vacuum. The result implies that we may ignore
the term $\rho _{ZPF}$, eq.$\left( \ref{ZPF}\right) ,$ which is equivalent
to putting $\rho _{ZPF}=0,$ for the calculation of the two-point correlation 
$C(r)$. That is from now on I should use eq.$\left( \ref{3.12}\right) $
ignoring the energy density operators $\hat{\rho}_{EM0}$ and $\hat{\rho}%
_{D0.}$

The calculation for free fields, that is to zeroth order in the interaction,
consists of substituting $\mid 0\rangle $ for $\mid vac\rangle $ in eq.$%
\left( \ref{2.11}\right) $ and neglecting the term $\hat{\rho}_{int}$ in eq.$%
\left( \ref{3.12}\right) .$ Thus the two-point correlation for free fields
(i.e. the unperturbed correlation) is the following 
\begin{eqnarray}
C_{0}(r) &=&\left\langle 0\left| \left[ \hat{\rho}_{EM}\left( \mathbf{r}%
_{1}\right) +\hat{\rho}_{D}\left( \mathbf{r}_{1}\right) \right] \left[ \hat{%
\rho}_{EM}\left( \mathbf{r}_{2}\right) +\hat{\rho}_{D}\left( \mathbf{r}%
_{2}\right) \right] \right| 0\right\rangle  \label{3.13} \\
&=&\left\langle 0\left| \hat{\rho}_{EM}\left( \mathbf{r}_{1}\right) \hat{\rho%
}_{EM}\left( \mathbf{r}_{2}\right) \right| 0\right\rangle +\left\langle
0\left| \hat{\rho}_{D}\left( \mathbf{r}_{1}\right) \hat{\rho}_{D}\left( 
\mathbf{r}_{2}\right) \right| 0\right\rangle ,  \nonumber
\end{eqnarray}
where the equality is a consequence of the fact that the cross terms $%
\left\langle 0\left| \hat{\rho}_{EM}\left( \mathbf{r}_{1}\right) \hat{\rho}%
_{D}\left( \mathbf{r}_{2}\right) \right| 0\right\rangle $ and $\left\langle
0\left| \hat{\rho}_{D}\left( \mathbf{r}_{1}\right) \hat{\rho}_{EM}\left( 
\mathbf{r}_{2}\right) \right| 0\right\rangle $ give no contribution. The
electromagnetic free field contribution becomes 
\begin{eqnarray*}
C_{EM0}(r) &=&\left\langle 0\left| \hat{\rho}_{EM}\left( \mathbf{r}%
_{1}\right) \hat{\rho}_{EM}\left( \mathbf{r}_{2}\right) \right|
0\right\rangle =\left\langle 0\left| \hat{\rho}_{EM2}\left( \mathbf{r}%
_{1}\right) \hat{\rho}_{EM2}\left( \mathbf{r}_{2}\right) \right|
0\right\rangle \\
&=&\frac{1}{16V^{2}}\left\langle 0\left| \sum_{\mathbf{k,\varepsilon }}\sum_{%
\mathbf{k}^{\prime }\mathbf{,\varepsilon }^{\prime }}K_{2}\hat{\alpha}_{%
\mathbf{k},\mathbf{\varepsilon }}\hat{\alpha}_{\mathbf{k}^{\prime },\mathbf{%
\varepsilon }^{\prime }}\sum_{\mathbf{k}^{\prime \prime }\mathbf{%
,\varepsilon }^{\prime \prime }}\sum_{\mathbf{k}^{\prime \prime \prime }%
\mathbf{,\varepsilon }^{\prime \prime \prime }}K_{2}^{*}\hat{\alpha}_{%
\mathbf{k}^{\prime \prime },\mathbf{\varepsilon }^{\prime \prime }}^{\dagger
}\hat{\alpha}_{\mathbf{k}^{\prime \prime \prime },\mathbf{\varepsilon }%
^{\prime \prime \prime }}^{\dagger }\right| 0\right\rangle ,
\end{eqnarray*}
where we have taken into account that the term involving $\hat{\rho}%
_{EM1}\left( \mathbf{r}\right) $ does not contribute (it annihilates the
unperturbed vacuum state $\mid 0\rangle ,$ see eq.$\left( \ref{2.3}\right) )$%
. Hence putting the operators in normal order, using the commutation rules,
and introducing an exponential convergence factor we get after some algebra 
\[
C_{EM0}(r)=\frac{1}{4V^{2}}\sum_{\mathbf{k,\varepsilon }}\sum_{\mathbf{k}%
^{\prime }\mathbf{,\varepsilon }^{\prime }}kk^{\prime }\left[ 1+\frac{%
\mathbf{k\cdot k}^{\prime }}{kk^{\prime }}\right] ^{2}\exp \left[ i\left( 
\mathbf{k}+\mathbf{k}^{\prime }\right) \mathbf{.r-}\varepsilon \left(
k+k^{\prime }\right) \right] , 
\]
where eqs.$\left( \ref{2.2b}\right) $ and $\left( \ref{2.2d}\right) $ have
been taken into account.

Substituting integrals for the sums this leads to 
\begin{eqnarray}
C_{EM0}(r) &=&\frac{1}{4\left( 2\pi \right) ^{6}}\int kd^{3}\mathbf{k}\int
k^{\prime }d^{3}\mathbf{k}^{\prime }\left[ 1+\frac{\mathbf{k\cdot k}^{\prime
}}{kk^{\prime }}\right] ^{2}  \nonumber \\
&&\times \exp \left[ i\left( \mathbf{k}+\mathbf{k}^{\prime }\right) \mathbf{%
.r-}\varepsilon \left( k+k^{\prime }\right) \right] .  \label{CEM}
\end{eqnarray}
The exponential convergence factor $\exp \left[ \mathbf{-}\varepsilon \left(
k+k^{\prime }\right) \right] $ is so chosen in order to get easy integrals,
analytical in some cases. It is plausible that the result would not depend
dramatically on the type of cut-off used, taking into acount that the
approximations involved in our calculation will allow only obtaining a rough
approximation. The details of the integals may be seen in the Appendix and
the result is 
\begin{equation}
C_{EM0}\left( r\right) =\frac{3r^{4}-10r^{2}\varepsilon ^{2}+3\varepsilon
^{4}}{\pi ^{4}(\varepsilon ^{2}+r^{2})^{6}}.  \label{EM0}
\end{equation}
The correlation $C_{EM0}\left( r\right) $ is positive for small $r$ as it
should, because for $r=0$ the correlation becomes the variance. That is a
fluctuation above (below) the average energy density is most probably close
to another one also above (below) it. For $r\in \left( \varepsilon
/3,3\varepsilon \right) $ the correlation is negative and it is again
positive for $r>3\varepsilon $.

A calculation of the two-point correlation of the free electron-positron
field starts from eq.$\left( \ref{2.6}\right) ,$ whence we get for $t=0$%
\begin{eqnarray*}
C_{D0}(r) &=&\left\langle 0\mid \hat{\rho}_{D}\left( \mathbf{r}_{1}\right) 
\hat{\rho}_{D}\left( \mathbf{r}_{2}\right) \mid 0\right\rangle =\left\langle
0\mid \hat{\rho}_{bd}\left( \mathbf{r}_{1}\right) \hat{\rho}_{bd}^{\dagger
}\left( \mathbf{r}_{2}\right) \mid 0\right\rangle \\
&=&\frac{1}{V^{2}}\sum_{\mathbf{pp}^{\prime }ss^{\prime }}\left| F\left( 
\mathbf{p,}s,\mathbf{p}^{\prime },s^{\prime }\right) \right| ^{2}\exp \left[
i\left( \mathbf{p+p}^{\prime }\right) \cdot \mathbf{r}\right] ,
\end{eqnarray*}
where we have taken into account that the terms $\rho _{b}\left( \mathbf{r}%
\right) $ and $\rho _{d}\left( \mathbf{r}\right) $ do not contribute because
they annihilate the state $\mid 0\rangle .$ Hence inserting the expression
of $\left| F\right| ^{2}$ eq.$\left( \ref{F2}\right) $, we obtain 
\begin{eqnarray}
C_{D0}(r) &=&\frac{1}{4V^{2}}\sum_{\mathbf{pp}^{\prime }}\left( E-E^{\prime
}\right) ^{2}\left[ 1+\frac{\mathbf{p\cdot q}-m^{2}}{EE^{\prime }}\right]
\exp \left[ i\left( \mathbf{p+p}^{\prime }\right) \cdot \mathbf{r}\right] 
\nonumber \\
&\rightarrow &\frac{1}{4(2\pi )^{6}}\int d^{3}\mathbf{p}\int d^{3}\mathbf{q}%
\left( E-E^{\prime }\right) ^{2}\left[ 1+\frac{\mathbf{p\cdot q}-m^{2}}{%
EE^{\prime }}\right]  \label{D} \\
&&\times \exp \left[ i\left( \mathbf{p+q}\right) \cdot \mathbf{r-}%
\varepsilon \left( p+q\right) \right] ,  \nonumber
\end{eqnarray}
after introducing a cut-off in the particles momenta. The integrals are
involved and they will not be calculated exactly. If $m<<\varepsilon ^{-1}$
we might approximate $E=\sqrt{p^{2}+m^{2}}\simeq p,E^{\prime }\simeq q$,
whence eq.$\left( \ref{D}\right) $ leads to 
\begin{equation}
C_{D0}\left( r\right) \simeq \frac{3r^{4}-10r^{2}\varepsilon
^{2}+3\varepsilon ^{4}}{2\pi ^{4}\left( r^{2}+\varepsilon ^{2}\right) ^{6}}%
+O\left( m^{2}\right) .  \label{DO}
\end{equation}
Details of the calculation may be seen in the Appendix.

The contributions of order $\alpha $ to the correlation are straightforward
although lengthy. They will not be reported in this paper.

\subsection{The relevance of metric fluctuations: A natural cut-off}

If spacetime is Minkowski then the two-point energy density correlation of
the vacuum fields may depend only on the interval between points that is,
with an obvious metric,

\begin{equation}
\left\langle vac\left| \hat{\rho}\left( \mathbf{r}_{1},t_{1}\right) \hat{\rho%
}\left( \mathbf{r}_{2},t_{2}\right) \right| vac\right\rangle =C\left( \sigma
\right) ,\sigma ^{2}\equiv \left( t_{2}-t_{1}\right) ^{2}-\left| \mathbf{r}%
_{1}-\mathbf{r}_{2}\right| ^{2}.  \label{3.1}
\end{equation}
This would be a consequence of the Lorentz invariance of the vacuum. Of
course the correlation might depend on whether the interval is timelike or
spacelike (in other words if $\sigma $ as defined above is real or
imaginary). However \textit{\ a Minkowski spacetime is not compatible with
the existence of fluctuations. }Therefore eq\textit{.}$\left( \ref{3.1}%
\right) $ is at most an approximation.

The proof of incompatibility is trivial and it follows. Let us consider a
Minkowski space and two arbitrary points, $A$ and $B$, with coordinates ($%
\mathbf{r}_{A},t_{A})$ and ($\mathbf{r}_{B},t_{B}),$ respectively. It is
always possible to find another point $C$ with coordinates ($\mathbf{r}%
_{C},t_{C})$ which is lightlike separated from $A$ and also from $B$. In
fact any point in the intersection of the light cones of $A$ and $B$ fulfils
that condition. Then eq.$\left( \ref{3.1}\right) $ implies that the vacuum
energy density in $C$ will be the same as in $A$ and the same as in $B$. But
the points $A$ and $B$ being arbitrary we conclude that the density will be
the same in all points of the Minkowski space. There would be no
fluctuations at all!

We may arrive at the same conclusion taking into account that if the
stress-energy tensor of the vacuum fields fluctuates, then the metric tensor
should also fluctuate by Einstein equation of general relativity. The
argument is clear in the context of classical physics and it should be taken
into account also in the quantum realm. In the calculation of the two-point
correlation eq.$\left( \ref{3.13}\right) $ we have used (implicitly) the
Minkowski \textit{classical} metric $diag\left( 1,1,1,-1\right) ,$ but we
should use a \textit{quantum} metric tensor operator $\hat{g}_{\mu \nu }(x)$%
, where $x$ represents the four coordinates of a point. That operator must
be related to the density operator, $\hat{\rho}(x),$ (more generally to the
stress-energy tensor operator) via a \textit{quantized Einstein equation}.
However a satisfactory quantum gravity theory is not yet available.
Therefore a correct calculation of the two-point correlation cannot be made.
With our present knowledge it is only possible to perform approximate
calculations that mix classical general relativity with quantum theory in
some form.

The standard method to deal with problems involving both quantum theory and
general relativity is the use of the ``semiclassical approximation''. It
consists of calculating quantum averages in a given spacetime (usually
Minkowski) and putting the averaged quantities on the right side of Einstein
equation, this treated as classical, in order to get the spacetime metric.
However the semiclassical approximation is not good enough for our purposes
in this paper. In fact its use leads to lossing two relevant properties of
the two-point density correlations of the vacuum fields. Firstly it does not
allow deriving the existence of a natural cut-off in the energies of the
particles associated to the vacuum fields, as is made in the following.
Secondly the approximation prevents calculating the long range effect on the
metric of the vacuum fluctuations, as will be discussed in section 4.

In order to go beyond the semiclassical approximation I will use a different
approach that consists of treating the metric fluctuations, actually
deriving from the quantum character of the metric, as if they were
associated to a classical statistical ensemble of metrics. That is I will
replace the correct, fully quantum, vacuum expectation by \textit{a
classical average over a statistical distribution of quantum vacuum
expectations}, each expectation calculated with a given classical metric.

More explicitly I will consider a probability distribution on an ensemble, $%
J,$ of metrics, all of them close (in some sense to be specified later) to a
reference Mikowski metric $diag(1,1,1,-1)$. I will use for every metric of
the ensemble $J$ the same (Cartesian) coordinate system as in Minkowski
space, but obviously a different metric tensor, $g_{\mu \nu }^{j},$ and a
different volume element. Our problem is to find a more accurate substitute
for the calculation made in the previous subsection, i. e. $C_{0}$ eq.$%
\left( \ref{3.13}\right) .$ Specifically the calculation in Minkowski space
required integrals involving dynamical variables, like momenta of the
particles, multiplied by position vectors, see for instance eqs.$\left( \ref
{CEM}\right) $ or $\left( \ref{D}\right) .$ We should find substitutes for
those integrals in any metric $g_{\mu \nu }^{j},$ and also we should find
the probability distribution in the ensemble of metrics. The calculation
would be involved, or impossible, and I shall make a simplification. I will
just consider a family of extremely simple metrics in order to get hints
about the main consequences of the metric fluctutations.

Let us consider a family of metrics with the form $diag(\gamma ,1,1,-1)$, $%
\gamma $ being a (positive) real number close to unity, and let us study the
change produced by that metric in a typical integral appearing in eq.$\left( 
\ref{CEM}\right) $ (where now we remove the cut-off, $\exp \left[ \mathbf{-}%
\varepsilon k\right] ,$because deriving the existence of a cut-off is just
the purpose of the present calculation ). For instance we will study the
change produced in the following integral 
\[
I\equiv \int kd^{3}\mathbf{k}\exp \left[ i\mathbf{k\cdot r}\right] =\int
dk_{x}\int dk_{y}\int dk_{z}\left| \mathbf{k}\right| \exp \left[ i\mathbf{%
k\cdot r}\right] , 
\]
where $\mathbf{k}$ is the momentum (or wavevector) of a virtual photon of
the vacuum. Actually the new metric, $diag(\gamma ,1,1,-1),$ corresponds to
a Minkowski space too, therefore it is plausible that a calculation in the
new metric would lead to the integral 
\[
I_{\gamma }\equiv \gamma \int dk_{x}\int dk_{y}\int dk_{z}\left| \mathbf{k}%
\right| \exp \left[ ik_{x}\gamma x+ik_{y}y+ik_{z}z\right] , 
\]
where the overall factor $\gamma $ takes the change of the volume element
into account. Now let us assume that the distribution of metrics in the
family corresponds to a Gaussian probability distribution, $f(\gamma )$,
where $\gamma $ is now taken as a random variable. That is 
\[
f(\gamma )d\gamma =\frac{1}{2\sqrt{\pi }\sigma }\exp \left( -\frac{(\gamma
-1)^{2}}{4\sigma ^{2}}\right) d\gamma , 
\]
(the probability $f(\gamma )$ is only approximately normalized because the
variable $\gamma $ cannot have negative values). The average of the
integrals $I_{\gamma }$ in the ensemble of metrics would be 
\[
\left\langle I_{\gamma }\right\rangle =\int_{0}^{\infty }I_{\gamma }f(\gamma
)d\gamma \simeq \int dk_{x}\int dk_{y}\int dk_{z}\left| \mathbf{k}\right|
\exp \left( i\mathbf{k\cdot r-}\sigma ^{2}k_{x}^{2}x^{2}\right) , 
\]
where with negligible error we have extended the $\gamma $ integration down
to $-\infty $ and approximated $\gamma $ by unity in the overall factor.

We could extend the family including other metrics like $diag(1,\gamma
,1,-1) $ or $diag(1,1,\gamma ,-1)$. The net result would be that the average
value of the integral $\left\langle I_{\gamma }\right\rangle $ should
exhibit a cut-off in the photon momenta. This illustrative calculation
supports the intuitive idea that randomness implies fuzzyness and fuzzyness
causes some average over photon momenta in the exponential $\exp \left( i%
\mathbf{k\cdot r}\right) .$ The fuzziness increases with the momentum $k$,
which amounts to an effective ultraviolet momenta cut-off. However neither
the calculation nor the intuition allow us finding the form of the cut-off.
Actually for the purpose of getting just a rough estimate of the effect of
the vacuum fluctuations the form of the cut-off is irrelevant. However an
estimate of the maximum allowed particle's momenta (say, an estimate of the
parameter $\sigma )$ is required. Finding such an esmate is difficult, but
arguments for its plausible order of mangitude follow.

It is common wisdom that there are quantum metric fluctuations at the Planck
scale. If these are the only relevant metric fluctuations then the momentum
cut-off should be of order the inverse of the Planck length, which would
imply an extremely large value for the two-point correlations (e.g those
calculated in eqs.$\left( \ref{CEM}\right) $ and $\left( \ref{D}\right) )$.
That is the energy of the vacuum fluctuations would be of order the vacuum
energy itself. It seems more plausible that the cut-off momentum is far
smaller, of order the typical particle scales. That is it should be of order
the inverse of a Compton wavelength, $h/mc$, $m$ being a typical (or
average) particles mass.

\section{Vacuum fluctuations and dark energy}

\subsection{The need of a cosmological constant in the Einstein equation}

It might be that the energy density of the different vacuum fields cancel to
each other, and similarly for the pressure, but our calculations within QED,
sections 2.2 and 2.3, do not support that assumption. A weaker assumption,
consistent with the desired properties of the vacuum, is that the whole set
of vacuum fields have a stress-energy tensor proportional to the metric
tensor. Furthermore our QED result suggests that this may be the case, but
the proportionality constant is divergent, or huge if we assume a natural
cut-off at the Planck scale. However this gives rise to the problem
commented in the introduction section, namely a big discrepancy between
theory and observations. A plausible solution to the problem is proposed in
the following.

We return to the discussion of section 1.2 about the relevance of the vacuum
fluctuations, but now in the realm of general relativity rather than
Newtonian gravity, although still within classical (not quantized) theories.
Einstein equation may be written 
\begin{equation}
R_{\mu \nu }-\frac{1}{2}g_{\mu \nu }R-\Lambda g_{\mu \nu }=-8\pi G\left(
T_{\mu \nu }^{matt}+T_{\mu \nu }^{vac}\right) ,  \label{Einst}
\end{equation}
(here I use the notation of the book by Weinberg\cite{Weinberg1}.) The
latter term of the left side involves the cosmological constant $\Lambda .$
On the right side the former term is due to matter and the latter to the
vacuum fields, both stress-energy tensors, $T_{\mu \nu }^{matt}$ and $T_{\mu
\nu }^{vac},$ depending on spacetime coordinates. The variation of the
vacuum tensor takes into account that vacuum fields fluctuate, see section
1-2. We will assume that these fluctuations are short ranged, as discussed
at the end of section 3.2.

Our proposed solution is that either the mean vacuum stress-energy tensor is
zero or it is precisely balanced by a cosmological constant, whilst the
fluctuations give rise to a long range effect able to explain dark energy.
If we pass from classical to quantum theory, then vacuum expectations should
be substituted for the averages over the volume $V$. Thus the main
assumptions in this paper may be stated as follows:

\begin{proposition}
\emph{The quantum expectation value of the stress-energy tensor operator due
to the vacuum fields is proportional to the expectation of the metric tensor
(or zero).}
\end{proposition}

\emph{\ }That is 
\begin{equation}
\left\langle T_{\mu \nu }^{vac}\right\rangle \equiv \langle vac\mid \hat{T}%
_{\mu \nu }^{vac}\mid vac\rangle =T^{vac}\langle vac\mid \hat{g}_{\mu \nu
}\mid vac\rangle ,  \label{Tl}
\end{equation}
where $T^{vac}$ is a constant and $\hat{g}_{\mu \nu }$ is the quantized
metric tensor operator. This implies that (the vacuum expectation of) the
stress-energy tensor of the combined vacuum fields has the form of a
cosmological constant in the Einstein equation, a hypothesis that is common
wisdom. That wisdom however suffers from a big difficulty, namely the said
cosmological constant would be huge, which leads us again to the problem
discussed in the introduction section, that is the disagreement between eqs.$%
\left( \ref{Planckdensity}\right) $ and $\left( \ref{darkdens}\right) .$ For
me the most plausible solution to the problem is the following assumption:

\begin{proposition}
\emph{There is a cosmological term in the Einstein equation that exactly
balances the effect of the mean vacuum stress-energy tensor. That is}
\end{proposition}

\begin{equation}
8\pi GT^{vac}=\Lambda .  \label{lambda}
\end{equation}

Assuming that the cancelation is exact we avoid any claim of conspiracy (see
section 1.2). Eq.$\left( \ref{lambda}\right) $ suggests writing the
quantized Einstein eq.$\left( \ref{Einst}\right) $ in the form 
\begin{equation}
\hat{G}_{\mu \nu }=-8\pi G\left( \hat{T}_{\mu \nu }^{matt}+\delta \hat{T}%
_{\mu \nu }^{vac}\right) ,\delta \hat{T}_{\mu \nu }^{vac}\equiv \hat{T}_{\mu
\nu }^{vac}-T^{vac}\hat{g}_{\mu \nu },  \label{EQ}
\end{equation}
where $\hat{G}_{\mu \nu }$, $\hat{T}_{\mu \nu }^{matt}$ and $\delta \hat{T}%
_{\mu \nu }^{vac}$ are tensor operators. The meaning of this equation is not
obvious because there is not yet a satisfactory quantum gravity theory and
therefore it is not known how to relate the metric tensor operator $\hat{g}%
_{\mu \nu }$ with the Einstein tensor operator $\hat{G}_{\mu \nu }.$ I shall
present in section 4.3 a procedure to give a sense to some approximation of
eq.$\left( \ref{EQ}\right) $ and to get sensible solutions. I will point out
that this equation departs from standard semiclassical approach to general
relativity, where the source term in a given quantum state should be the
expectation value of the relevant field operators in that state. In fact,
the semiclassical alternative to eq.$\left( \ref{EQ}\right) $ would be 
\begin{equation}
G_{\mu \nu }=-8\pi G\langle vac\mid \hat{T}_{\mu \nu }^{matt}+\delta \hat{T}%
_{\mu \nu }^{vac}\mid vac\rangle ,  \label{deltaT}
\end{equation}
but the term $\langle vac\mid \delta \hat{T}_{\mu \nu }^{vac}\mid vac\rangle 
$ has been assumed nil, see eq.$\left( \ref{Tl}\right) .$ Consequently the
effect of the vacuum fluctuations would be lost in a standard semiclassical
treatment.

\subsection{The gravity of vacuum fluctuations. A toy model}

In order to study the spacetime curvature produced by density and pressure
fluctuations we must start from Einstein eq.$\left( \ref{Einst}\right) $,
which leads to the quantized Einstein eq.$\left( \ref{EQ}\right) $ with our
assumptions$.$ Ignoring the effect of matter, it simplifies to

\begin{equation}
\hat{G}_{\mu \nu }=-8\pi G\delta \hat{T}_{\mu \nu }^{vac},  \label{EQ1}
\end{equation}
where the Einstein tensor operator $\hat{G}_{\mu \nu }$ is a functional of
the metric tensor operator $\hat{g}_{\mu \nu }$. The functional $\hat{G}%
_{\mu \nu }\left[ \hat{g}_{\lambda \sigma }\right] $ would contain $\hat{g}%
_{\lambda \sigma }$ and its first and second derivatives with respect to the
coordinates, but we do not know what is the exact relation, something that
should be derived from a quantum gravity theory. As a consequence eq.$\left( 
\ref{EQ1}\right) $ is not well defined. Nevertheless for the study of vacuum
fluctuations an approximate quantum equation may be substituted for eq.$%
\left( \ref{EQ1}\right) ,$ and sensible solutions found, as shown in the
next section.

Before presenting that solution, I will work an extremely simplified model
where the two-point density correlations depend on a single parameter, the
radial coordinate. Of course the model is far from realistic, but it
provides hints about the actual consequences of the vacuum fluctuations. I
consider a static problem with spherical symmetry in standard (curvature)
coordinates $\left\{ r,\theta ,\phi ,t\right\} $ with some stress-energy
tensor whose components fulfil 
\begin{equation}
\left\langle vac\left| (\delta \hat{T}^{vac})_{\mu }^{\nu }\right|
vac\right\rangle =0,  \label{expectrop}
\end{equation}
as a consequence of eq.$\left( \ref{Tl}\right) .$ Using the metric 
\begin{equation}
d\sigma ^{2}=A\left( r\right) dr^{2}+r^{2}\left[ d\theta ^{2}+\sin
^{2}\theta d\phi ^{2}\right] -B\left( r\right) dt^{2},  \label{metric}
\end{equation}
one of the components of the classical Einstein's equation is\cite{Weinberg1}
\begin{equation}
\frac{d}{dr}\left( \frac{r}{A}\right) =1-8\pi Gr^{2}\rho \left( r\right) ,
\label{Einst3}
\end{equation}
with the initial condition $A\left( 0\right) =1.$ We shall work in quantized
general relativity. Although no satisfactory quantum gravity theory exists
yet, it is plausible that the relations between the quantum operators are
the same as the classical ones whenever there is no problem with the
non-commutativity of the operators. Therefore the quantized counterpart of
eq.$\left( \ref{Einst3}\right) $ relevant for our problem would be 
\begin{equation}
\frac{d}{dr}\left( \frac{r}{\hat{A}}\right) =1-8\pi Gr^{2}\delta \hat{\rho}%
\left( r\right) ,  \label{Einst6}
\end{equation}
where $8\pi \delta \hat{\rho}$ is the component of the stress-energy tensor, 
$(\delta \hat{T}^{vac})_{\mu }^{\nu },$ derived from the vacuum
fluctuations. I stress that the average tensor due to the vacuum fields does
not appear in eq.$\left( \ref{6}\right) $ because I have assumed that it is
exactly canceled by a cosmological constant. However for notational
simplicity I will substitute $\hat{\rho}$ for $\delta \hat{\rho}$ from now
on, but take into account that we will have 
\begin{equation}
\left\langle vac\left| \hat{\rho}\right| vac\right\rangle =0.  \label{Einst7}
\end{equation}
The exact solution of this operator equation is trivial and it may be
written in closed form, that is 
\begin{equation}
\hat{A}=\left[ 1-\frac{8\pi G}{r}\int_{0}^{r}x^{2}\hat{\rho}\left( x\right)
dx\right] ^{-1}.  \label{Einst4}
\end{equation}
Now we assume that Newton constant $G$ is small (e. g. in comparison with $c 
%TCIMACRO{\UNICODE[m]{0x127}}
%BeginExpansion
\rlap{\protect\rule[1.1ex]{.325em}{.1ex}}h%
%EndExpansion
/m^{2}$, $m$ being a typical particle mass). Thus we expand eq.$\left( \ref
{Einst4}\right) $ in powers of $G$ giving 
\begin{eqnarray}
\hat{A} &=&1+\frac{8\pi G}{r}\int_{0}^{r}x^{2}\hat{\rho}\left( x\right) dx 
\nonumber \\
&&+\frac{64\pi ^{2}G^{2}}{r^{2}}\int_{0}^{r}x^{2}\hat{\rho}\left( x\right)
dx\int_{0}^{r}y^{2}\hat{\rho}\left( y\right) dy+O\left( G^{3}\right) .
\label{A}
\end{eqnarray}
The quantized densities at different points may not commute, that is we
might have 
\[
\left[ \hat{\rho}\left( x\right) ,\hat{\rho}\left( y\right) \right] \neq 0, 
\]
but this is irrelevant in eq.$\left( \ref{A}\right) $ because the latter
term is itself the square of an operator. The interesting quantity is the
vacuum expectation of the metric, which would provide the observable
space-time curvature. The expectation of the above equation leads to 
\begin{equation}
\left\langle vac\left| \hat{A}\right| vac\right\rangle \simeq 1+\frac{64\pi
^{2}G^{2}}{r^{2}}\int_{0}^{r}x^{2}dx\int_{0}^{r}y^{2}dy\left\langle
vac\left| \hat{\rho}\left( x\right) \hat{\rho}\left( y\right) \right|
vac\right\rangle ,  \label{A1}
\end{equation}
where eq.$\left( \ref{expectrop}\right) $ has been taken into account. As a
conclusion we have obtained the expectation value of one of the metric
components in terms of the two-point correlation of the energy density, and 
\textit{shown that the metric component may have long range (of order }$r$%
\textit{) deviations from Minkowski metric, in spite of the vacuum
expectation of the density being zero}, see eq.$\left( \ref{Einst7}\right) .$
I stress that this result could not be derived from the standard
``semiclassical approximation'' to general relativity. In that approximation
we should write, instead eq.$\left( \ref{Einst6}\right) ,$ the following 
\[
\frac{d}{dr}\left( \frac{r}{\left\langle vac\left| \hat{A}\right|
vac\right\rangle }\right) =1-8\pi Gr^{2}\left\langle vac\left| \delta \hat{%
\rho}\left( r\right) \right| vac\right\rangle =1, 
\]
whence we would get 
\[
\left\langle vac\left| \hat{A}\right| vac\right\rangle =1, 
\]
that is the Minkowski value.

We may search for an effective (classical) density, $\rho _{eff}\left(
r\right) ,$ providing the same spacetime curvature as the vacuum expectation
of the metric component, eq.$\left( \ref{A1}\right) $. It would fulfil 
\[
\left\langle vac\left| \hat{A}\right| vac\right\rangle =1+\frac{8\pi G}{r}%
\int_{0}^{r}x^{2}\rho _{eff}\left( x\right) dx+O\left( G^{2}\right) , 
\]
that after inserting it in eq.$\left( \ref{A1}\right) $ and calculating the
derivative leads to 
\begin{equation}
\rho _{eff}\left( r\right) \simeq \frac{8\pi G}{r^{2}}\frac{d}{dr}\left[ 
\frac{1}{r}\int_{0}^{r}x^{2}dx\int_{0}^{r}y^{2}dy\left\langle vac\left| \hat{%
\rho}\left( x\right) \hat{\rho}\left( y\right) \right| vac\right\rangle
\right] .  \label{roefr}
\end{equation}
This result emphasizes the main consequence to be derived from our toy
model, namely that the vacuum fluctuations produce a metric equivalent to
the metric derived from some effective (fictitious) density.

The second component of the classical Einstein equation cannot be quantized
as easily as eq.$\left( \ref{Einst3}\right) ,$ so that we may follow a
different route in order to get the quantized component of the metric, $\hat{%
B}(r)$, see eq.$\left( \ref{metric}\right) .$ We start from the well known
classical solution 
\begin{equation}
B(r)=\exp \left[ G\int_{0}^{r}\frac{2M(x)+8\pi x^{3}P\left( x\right) }{%
x-2GM(x)}dx\right] ,  \label{B}
\end{equation}
where 
\[
M(x)\equiv \int_{0}^{x}4\pi y^{2}\rho \left( y\right) dy, 
\]
$P$ being the pressure (more correctly the component $\delta T_{1}^{1}$ of
the stress-energy tensor eq.$\left( \ref{EQ1}\right) $ divided by $-8\pi $).
Expanding eq.$\left( \ref{B}\right) $ to second order in $G$ we get 
\begin{eqnarray*}
B(r) &=&1+G\int_{0}^{r}\frac{2M(x)+8\pi x^{3}P\left( x\right) }{x}dx+\frac{1%
}{2}G^{2}\left[ \int_{0}^{r}\frac{2M(x)+8\pi x^{3}P\left( x\right) }{x}%
dx\right] ^{2} \\
&&+2G^{2}\int_{0}^{r}\frac{[2M(x)+8\pi x^{3}P\left( x\right) ]M(x)}{x^{2}}%
dx+O\left( G^{3}\right) .
\end{eqnarray*}
In order to quantize this expression we should substitute operators for the
dynamical variables. However there is a difficulty due to the possible
noncommutativity of $\hat{P}\left( x\right) $ and $\hat{M}(x).$ I propose
the following rule:

\textit{For the quantization of the product of \textbf{two} components of
the stress-energy tensor we should use the symetrized product. }

This leads to 
\begin{eqnarray*}
P\left( x\right) M(x) &=&\int_{0}^{x}4\pi y^{2}dyP\left( x\right) \rho (y) \\
&\rightarrow &\int_{0}^{x}2\pi y^{2}dy\left[ \hat{P}\left( x\right) \hat{\rho%
}(y)+\hat{\rho}(y)\hat{P}\left( x\right) \right] \\
&=&\frac{1}{2}\left[ \hat{P}\left( x\right) \hat{M}(x)+\hat{\rho}(y)\hat{M}%
\left( x\right) \right] .
\end{eqnarray*}
Then the vacuum expectation of the quantized metric component $\hat{B}(r)$
becomes 
\begin{eqnarray}
\left\langle vac\left| \hat{B}\right| vac\right\rangle
&=&1+4G^{2}\int_{0}^{r}\frac{\left\langle vac\left| \left[ \hat{M}(x)\right]
^{2}\right| vac\right\rangle }{x^{2}}dx\smallskip  \label{P} \\
&&+\frac{1}{2}G^{2}\left\langle vac\left| \left[ \int_{0}^{r}\frac{2\hat{M}%
(x)+8\pi x^{3}\hat{P}\left( x\right) }{x}dx\right] ^{2}\right|
vac\right\rangle  \nonumber \\
&&+8\pi G^{2}\int_{0}^{r}\left\langle vac\left| \left[ \hat{P}\left(
x\right) \hat{M}\left( x\right) +\hat{M}\left( x\right) \hat{P}\left(
x\right) \right] \right| vac\right\rangle xdx.  \nonumber
\end{eqnarray}
We may introduce an effective pressure, $P_{eff}\left( r\right) ,$ such that 
\begin{eqnarray}
\left\langle vac\left| \hat{B}\right| vac\right\rangle &=&1+\exp \left[
G\int_{0}^{r}\frac{2M_{eff}(x)+8\pi x^{3}P_{eff}\left( x\right) }{%
x-2GM_{eff}(x)}dx\right]  \nonumber \\
&=&8\pi G\int_{0}^{r}\left[ P_{eff}\left( x\right) +\ln \left( \frac{r}{x}%
\right) \rho _{eff}\left( x\right) \right] x^{2}dx+O(G^{2}).  \label{pefr}
\end{eqnarray}
It is straightforward to get $P_{eff}(r)$ equating this with eq.$\left( \ref
{P}\right) ,$ but I will not write the result.

\subsection{Approximate solution in four dimensions}

In order to devise a method for the solution of eq.$\left( \ref{EQ}\right) ,$
ignoring the effect of matter, I propose the following generalization of the
method used in the previous toy model.

We should write the metric tensor, $g_{jk},$ as an expansion in powers of
the Newton constant, $G$, that may be taken as a small parameter whence the
deviation from the Minkowski metric is assumed small. Therefore we write 
\[
g_{jk}=g_{jk}^{(0}+Gg_{jk}^{(1}+G^{2}g_{jk}^{(2}... 
\]
Thus the zeroth order approximation of eq.$\left( \ref{Einst2}\right) $
becomes 
\[
G_{\mu \nu }\left[ g_{jk}^{(0}\right] =0, 
\]
whence $g_{jk}^{(0}$ is the metric tensor of Minkowski space. The first
order approximation will be 
\[
-8\pi \delta T_{\mu \nu }^{vac0}\left( x\right) =\left( \frac{\delta G_{\mu
\nu }\left[ g_{jk}\right] }{\delta g_{jk}}\right) g_{jk}^{(1}, 
\]
where the functional derivative should be taken at $g_{jk}=g_{jk}^{(0}$. The
point is that this equation is \textit{linear} in the unknown, therefore
relatively easy to solve. The second order approximation would involve the
second functional derivative, giving another linear equation, and so on. In
summary we reduce the Einstein nonlinear equation to an infinite set of
linear ones. Of course this is only valid because we are assuming here that
the metric is close to Minkowski, with small fluctuations superimposed on
it. As explained below the second order is sufficient for the study of the
vacuum fluctuations.

An advantage of the approximate method to solve Einstein eq.$\left( \ref
{Einst2}\right) $ is that it may be used, with plausible assumptions, for a
treatment in \emph{quantized} general relativity. In the quantized theory
the tensors $T_{\mu \nu }^{vac}$ and $g_{jk}$ become operators, but their
relation cannot be obtained from the classical (Einstein) equation due to
the possible non-commutativity. However, the approximate relations between
the metric and the stress-energy tensors, involved in the calculation up to
second order in $G,$ are either linear or they contain products of no more
than two components of the stress-energy operator. In the linear relations
there is no problem of commutativity and in relations involving products of
two operators it is plausible to use the symmetrized product. That is if $%
T_{\mu \nu }^{vac}\left( x\right) T_{\lambda \sigma }^{vac}\left( x^{\prime
}\right) $ is a classical product, the quantum counterpart would be 
\[
1/2\left[ \hat{T}_{\mu \nu }^{vac}\left( x\right) \hat{T}_{\lambda \sigma
}^{vac}\left( x^{\prime }\right) +\hat{T}_{\lambda \sigma }^{vac}\left(
x^{\prime }\right) \hat{T}_{\mu \nu }^{vac}\left( x\right) \right] , 
\]
whee $x$ represents the four coordinates of a spacetime point$.$

The conclusions of this section are the following:

1) It is possible to quantize the Einstein equation of general relativity in
the post-Newtonian approximation, that is neglecting terms of order higher
than $G^{2}$ provided we assume that the quantum counterpart of the product
of two components of the stress-energy tensor gives a symmetrized product of
operators.

2) The vacuum expectation of the metric induced by the quantum vacuum
fluctuations is the same as the one due to a classical effective
stress-energy tensor, that is an effective mass density, $\rho _{eff}$, and
pressure, $P_{eff}$ if the spacetime is Minkowski (at least with good enough
approximation).

3) The effective mass density is given by Newton constant, $G$, times a
quantity which may depend only on the masses of the particles involved (i.
e. the electron mass if we study just QED effects) and the universal
constants.

\subsection{An estimate of the dark energy density}

In the previous sections of this paper I have sketched a research program
that in principle would allow calculating the properties of dark energy in
terms of the fundamental quantum fields. I exclude the possibility that dark
energy may be identified with the mean stress-energy of the vacuum fields,
which would give an unplausibly huge value. In fact I assume that a
cosmological constant exists in Einstein equation that exactly balances that
mean field energy, but I suppose that the cosmological constant cannot
balance the energy of the vacuum fluctuations. Thus a calculation of the
dark energy properties should start from the study of the two-point
correlations of the fluctuations of the vacuum stress-tensor. It is shown
that the result would be finite due to metric fluctuations (see section 3).
The vacuum expectation of the two-point correlation gives rise to a long
range spacetime curvature as suggested in subsection 4.3, which would allow
in principle getting the effective stress-energy tensor. It is
straightforward to get a classical effective stress-energy tensor providing
the same curvature as the two-point quantum correlation. Finally we should
identify that effective mass density with the density, eq.$\left( \ref
{darkdens}\right) ,$ of the dark energy thus getting the properties of dark
energy from fundamental quantum fields.

Actually a calculation along the lines of the previous subsection, involving
a simplified static model (i. e. a model in 3 rather than 4 dimensions), has
been worked elsewhere\cite{Santos}. The model is more realistic than the one
of section 4.2 in the sense that it does not assume spherical symmetry. That
is the two-point correlations of the stress-energy tensors depend on the
spatial distance between the points, rather than on the difference between
the radial coordinates as in the model of section 4.2. However the main
conclusion is similar as in ours eqs.$\left( \ref{roefr}\right) $ and $%
\left( \ref{pefr}\right) .$ Namely a stress-energy tensor operator with nil
vacuum expectation gives rise to a long range spacetime curvature equivalent
to the one produced by an effective (classical) stress-energy tensor.That
tensor is proportional to some integral of the two-point correlation $C(r).$
In particular with some plausible assumptions the said model\cite{Santos}
gives the following effective density, $\rho _{eff},$ and pressure, $P_{eff}$%
, 
\begin{equation}
\rho _{eff}=-P_{eff}\sim KG\int_{0}^{\infty }C(r)rdr,  \label{ASS}
\end{equation}
where the numerical constant $K$ in front of the integral is of order unity.
It is interesting that $\rho _{eff}$ so obtained roughly equals the
gravitational energy of the two-point correlations if calculated with
Newtonian gravity, eq.$\left( \ref{0.1}\right) $. If the vacuum fluctuations
are the cause of the dark energy we should identify $\rho _{eff}$ in eq.$%
\left( \ref{ASS}\right) $ with $\rho _{DM}$ in eq.$\left( \ref{darkdens}%
\right) $ whence we get the following estimate for the QED contribution to
the dark energy density 
\begin{equation}
\rho _{DM}\sim KG\int_{0}^{\infty }C\left( r\right) rdr\sim G\varepsilon
^{-6}\left[ 1+O(m^{2}\varepsilon ^{2})\right] ,  \label{roeff}
\end{equation}
where $C\left( r\right) $ is identified our calculated value in eqs.$\left( 
\ref{EM0}\right) $ and $\left( \ref{DO}\right) $.

This is the QED contribution to the dark energy density, that is the
contribution of the electromagnetic and electron-positron vacuum fields
(calculated to zeroth order in the perturbation). It is plausible that the
contribution of all quantum fields would be of order eq.$\left( \ref{roeff}%
\right) $ times a few tens. In any case the main obstacle for a valid
estimate of the dark energy density is the difficulty to calculate the
cut-off derived from the metric fluctuations, as discussed in subsection
3.2. Before solving this problem it is not worth to refine the calculations.

A guess about the value of the said cut-off may be taken as follows. The
parameter $\varepsilon $ was introduced in section 3.1 as the inverse of the
maximum particle's momentum and we have argued in section 3.2 that a
plausible estimate should be of order of some average particle mass, $m,$
times $c$. Thus we get 
\[
\rho _{DM}=\rho _{eff}\sim \frac{Gm^{6}c^{2}}{%TCIMACRO{\UNICODE[m]{0x127}}
%BeginExpansion
\rlap{\protect\rule[1.1ex]{.325em}{.1ex}}h%
%EndExpansion
^{4}}, 
\]
where we have written explicitly the velocity of light and Planck\'{}s
constant for clarity. That quantity agrees with the observed value, eq.$%
\left( \ref{darkdens}\right) ,$ if we put $m$ about 80 times the electron
mass, which is not unplausible.

\section{Excited vacuum states and dark matter}

In section 2.4 we have studied the physical vacuum state, defined as the
ground state of the total Hamiltonian, $H,$ of the interacting quantum
fields, that is the eigenstate with the smallest eigenvalue. $H$ should
include all quantum fields of nature toghether with their interactions, but
in this paper we have used the QED Hamiltonian as an example. It is a
straightforward prediction of quantum theory that other eigenstates of $H$
are possible in addition to the ground state. Amongst these states there is
a particular class, studied in section 2.4, that I have named \emph{excited
vacuum states (EVS). }We may ask whether those states may be observed
somewhere and I propose that they may be candidates for dark matter.

The relevant property is that \emph{EVS} consist of coupled particles
possessing the same quantum numbers of the vacuum, that is no charges
(either electric, baryonic, leptonic, etc.), and also no net momentum or
angular momentum. As a consequence they would not interact via
electromagnetic, weak or strong nuclear forces. However such states possess
energy above the vacuum, see eq.$\left( \ref{Ej}\right) $. In section 2.5 I
have assumed that the total energy of the vacuum state is exactely balanced
by a cosmological constant, but the \emph{EVS} possess energy above the
vacuum and therefore these states should interact gravitationally.

In section 2.4 we studied \emph{EVS} in a Minkowski space, that is in
absence of graviational fields, therefore having traslational and rotational
invariance. However it is plausible that in the presence of gravitational
fields of baryonic matter the states are modified, suffering the atraction
of large masses. A more detailed study would be needed in order to know the
possible gravitational modification of the \emph{EVS}. In the meantime I
conjecture that the effect would be relevant only for extended gravitational
fields like those of galaxies and clusters. In any case the \emph{EVS} have
properties not too different from those usually adscribed to the
hypothetical constituents of dark matter, that is they are cold dark matter,
as they possess energy but no pressure (or very small). A more detailed
study of the possibility that dark matter consists of \emph{EVS} will not be
made in this paper, but I believe that such study is worth to be made.

\section{Conclusions}

The energy distribution in the quantum vacuum has been studied for the
particular case of quantum electrodynamics (QED). It has been shown that the
stress-energy tensor of the combined electromagnetic and electron-positron
fields is roughly proportional to the metric tensor, but the proportionality
constant diverges. As a plausible solution it is proposed that there is a
cosmological constant in Einsein equation that exactly balances the
stress-energy of the quantum vacuum. Arguments have been given showing that
vacuum fluctuations give rise to a metric able to explain dark energy.
Finally it has been suggested that dark matter may be an effect of vacuum
excitations.

\section{Appendix. Two-point correlation to zeroth order}

\textbf{The free electromagnetic field}

We shall solve the integrals in eq.$\left( \ref{CEM}\right) ,$ that is

\begin{eqnarray*}
C_{EM}(\mathbf{r}) &=&\frac{1}{256\pi ^{6}}\int d^{3}\mathbf{k}\int d^{3}%
\mathbf{k}^{\prime }kk^{\prime }\left[ 1+\frac{\mathbf{k\cdot k}^{\prime }}{%
kk^{\prime }}\right] ^{2} \\
&&\times \exp \left[ i\left( \mathbf{k+k}^{\prime }\right) \cdot \mathbf{r}%
-\varepsilon k-\varepsilon k^{\prime }\right] .
\end{eqnarray*}
In spherical coordinates with polar angle in the direction of $\mathbf{r}$
we get 
\begin{eqnarray*}
C_{EM}(\mathbf{r}) &=&\frac{1}{256\pi ^{6}}\int_{0}^{\infty
}k^{3}dk\int_{0}^{\infty }k^{\prime 3}dk^{\prime }\int_{-1}^{1}d\left( \cos
\theta \right) \int_{-1}^{1}d\left( \cos \theta ^{\prime }\right) \\
&&\times \int_{0}^{2\pi }d\phi \int_{0}^{2\pi }d\phi ^{\prime }\exp \left(
ikr\cos \theta -\varepsilon k\right) \exp \left( ik^{\prime }r\cos \theta
^{\prime }-\varepsilon k^{\prime }\right) \\
&&\times \left[ 1+\sin \theta \cos \phi \sin \theta ^{\prime }\cos \phi
^{\prime }+\sin \theta \sin \phi \sin \theta ^{\prime }\sin \phi ^{\prime
}+\cos \theta \cos \theta ^{\prime }\right] ^{2}.
\end{eqnarray*}
Integration of the angles $\phi $ and $\phi ^{\prime }$ leads to 
\begin{eqnarray*}
C_{EM}(\mathbf{r}) &=&\frac{1}{128\pi ^{4}}\int_{0}^{\infty
}k^{3}dk\int_{0}^{\infty }k^{\prime 3}dk^{\prime
}\int_{-1}^{1}du\int_{-1}^{1}du^{\prime } \\
&&\times \left[ 3+3u^{2}u^{\prime 2}-u^{2}-u^{\prime 2}+4uu^{\prime }\right]
\exp \left( ikru-\varepsilon k\right) \exp \left( ik^{\prime }ru^{\prime
}-\varepsilon k^{\prime }\right) ,\text{ }
\end{eqnarray*}
where $r\equiv \left| \mathbf{r}\right| ,$ $u\equiv \cos \theta ,u^{\prime
}\equiv \cos \theta ^{\prime }.$ In terms of the integrals to be defined
below we get 
\begin{eqnarray*}
C_{EM}(\mathbf{r}) &=&\frac{1}{32\pi ^{4}}\left[
3I_{3}^{2}+3I_{3uu}^{2}-2I_{3}I_{3uu}+4I_{3u}^{2}\right] \\
&=&\frac{3r^{4}-10r^{2}\varepsilon ^{2}+3\varepsilon ^{4}}{\pi ^{4}\left(
r^{2}+\varepsilon ^{2}\right) ^{6}}
\end{eqnarray*}

\textbf{The free electron-positron field}

We shall start from eq.$\left( \ref{D}\right) ,$ that is

\begin{eqnarray}
C_{D}\left( \mathbf{r}\right) &=&\frac{1}{256\pi ^{6}}\int d^{3}\mathbf{p}%
\int d^{3}\mathbf{q}\left( E-E^{\prime }\right) ^{2}\left[ 1+\frac{\mathbf{%
p\cdot q}-m^{2}}{EE^{\prime }}\right]  \nonumber \\
&&\times \exp \left[ i\left( \mathbf{p+q}\right) \cdot \mathbf{r-}%
\varepsilon \left( p+q\right) \right] ,  \label{10}
\end{eqnarray}
that in polar coordinates becomes 
\begin{eqnarray*}
C_{D}\left( \mathbf{r}\right) &=&\frac{1}{64\pi ^{4}}\int_{0}^{\infty
}p^{2}dp\int_{0}^{\infty }q^{2}dq\int_{-1}^{1}du\int_{-1}^{1}du^{\prime } \\
&&\times \left( E-E^{\prime }\right) ^{2}\left[ 1+\frac{pquu^{\prime }-m^{2}%
}{EE^{\prime }}\right] \\
&&\times \exp \left[ i\left( pu+qu^{\prime }\right) r\mathbf{-}\varepsilon
\left( p+q\right) \right]
\end{eqnarray*}
If we approximate 
\[
E\simeq p+\frac{m^{2}}{2p},E^{\prime }\simeq p+\frac{m^{2}}{2p}, 
\]
valid for $m<<\varepsilon ^{-1},$ we get to order $O(m^{2})$%
\begin{eqnarray*}
&&\left( E-E^{\prime }\right) ^{2}\left[ 1+\frac{pquu^{\prime }-m^{2}}{%
EE^{\prime }}\right] \\
&\simeq &\left( p-q\right) ^{2}(1+uu^{\prime })-2m^{2}\frac{\left(
p-q\right) ^{2}}{pq}+\frac{m^{2}}{2}uu^{\prime }\frac{\left( p-q\right) ^{4}%
}{p^{2}q^{2}}.
\end{eqnarray*}
The integrals that appear in the terms of order $m^{2}$ are lengthy,
although straightforward, and they will not be reporte here. The term of
zeroth order is 
\begin{eqnarray*}
C_{D0}\left( r\right) &\simeq &\frac{1}{8\pi ^{4}}\left[
I_{2}I_{4}-I_{3}^{2}+I_{2u}I_{4u}-I_{3u}^{2}\right] +O(m^{2}) \\
&=&\frac{3r^{4}-10r^{2}\varepsilon ^{2}+3\varepsilon ^{4}}{2\pi ^{4}\left(
r^{2}+\varepsilon ^{2}\right) ^{6}}+O(m^{2}).
\end{eqnarray*}
\noindent

\textbf{The integrals }

The\textrm{\ }required integrals are\textrm{\ } 
\[
I_{n}\equiv \frac{1}{2}\int_{0}^{\infty }k^{n}dk\exp \left( -\varepsilon
k\right) \int_{-1}^{1}du\cos \left( kru\right) , 
\]
\[
I_{nu}\equiv \frac{i}{2}\int_{0}^{\infty }k^{n}dk\exp \left( -\varepsilon
k\right) \int_{-1}^{1}udu\sin \left( kru\right) , 
\]
\[
I_{nuu}\equiv \frac{1}{2}\int_{0}^{\infty }k^{n}dk\exp \left( -\varepsilon
k\right) \int_{-1}^{1}u^{2}du\cos \left( kru\right) . 
\]
The results for the relevant integrals are as follows 
\[
I_{2}=\frac{2\varepsilon }{(\varepsilon ^{2}+r^{2})^{2}},I_{3}=\frac{%
6\varepsilon ^{2}-2r^{2}}{(\varepsilon ^{2}+r^{2})^{3}},I_{4}=\frac{%
24\varepsilon \left( \varepsilon ^{2}-r^{2}\right) }{(\varepsilon
^{2}+r^{2})^{4}}, 
\]
\[
I_{2u}=\frac{2ir}{(\varepsilon ^{2}+r^{2})^{2}},I_{3u}=\frac{8ir\varepsilon 
}{(\varepsilon ^{2}+r^{2})^{3}}, 
\]
\[
I_{4u}=-8ir\frac{r^{2}-5\varepsilon ^{2}}{(\varepsilon ^{2}+r^{2})^{4}}%
,I_{3uu}=\frac{2\varepsilon ^{2}-6r^{2}}{(\varepsilon ^{2}+r^{2})^{3}}. 
\]

\end{document}